\documentstyle[preprint,aps]{revtex}
\begin{document}
\draft
\title{Phenomenological Renormalization Group Methods}

\author{ J. A. Plascak }
\address{ Departamento de F\'\i sica, Instituto de Ci\^encias Exatas,\\
 Universidade Federal de Minas Gerais, C.P. 702, \\
 30161-123 Belo Horizonte, Brazil \\
 e-mail pla@fisica.ufmg.br}
\author{W. Figueiredo and B. C. S. Grandi}
\address{Departamento de Fisica, \\
Universidade Federal de Santa Catarina\\
88040-900 Florian\'opolis, Brazil \\
e-mail wagner@fisica.ufsc.br and bartirag@fisica.ufsc.br }
\date{\today}
\maketitle

\begin{abstract}

Some renormalization group approaches have been proposed during
the last few years which are close in spirit to the Nightingale 
phenomenological procedure. In essence, by exploiting the finite
size scaling hypothesis, the approximate critical behavior of the model on 
infinite lattice is obtained through the exact computation of some thermal
quantities of the model on finite clusters. In this work some 
of these methods are reviewed, namely the mean field renormalization
group, the effective field renormalization group and the finite
size scaling renormalization group procedures. Although
special emphasis is given to the mean field renormalization group (since
it has been, up to now, much more applied an extended to study a wide
variety of different systems) a discussion of their potentialities 
and interrelations to other methods is also addressed.

\end{abstract} 

\narrowtext

\noindent {\bf Contents}
\vskip 0.1truecm
{I. } INTRODUCTION
\hfil ................................................ 04\break
{II. } PHENOMENOLOGICAL MFRG
\hfil ................................................ 13\break
\null~~~~II.A MFRG
\hfil ................................................ 13\break
\null~~~~II.B SURFACE-BULK MFRG
\hfil ................................................ 21\break
\null~~~~II.C EXTENDED MFRG
\hfil ................................................ 24\break
\null~~~~~~~~ II.C.1 One Order Parameter
\hfil ................................................ 25\break
\null~~~~~~~~ II.C.2 Two Uncoupled Order Parameters
\hfil ................................................ 28\break
\null~~~~~~~~ II.C.3 Two Coupled Order Parameters
\hfil ................................................ 29\break
\null~~~~II.D DYNAMIC MFRG
\hfil ................................................ 31\break
{III. } EXACT RESULTS FROM MFRG
\hfil ................................................ 36\break
{IV. }  STATIC PROBLEMS APPLICATIONS
\hfil ................................................ 37\break
{V. } DYNAMIC PROBLEMS APPLICATIONS
\hfil ................................................ 38\break
{VI. } RELATED PHENOMENOLOGICAL RG
\hfil ................................................ 42\break
\null~~~~VI.A PHENOMENOLOGICAL EFRG
\hfil ................................................ 42\break
\null~~~~VI.B PHENOMENOLOGICAL FSSRG
\hfil ................................................ 46\break
{VII. } FINAL REMARKS  \hfil 
      ................................................ 50\break
References \hfil
      ................................................ 61\break

\eject

\section{Introduction}

The renormalization group formalism introduced by Wilson in the early 70's 
\cite{wilson1,wilson2} is by now one of the basic strategies to solve
fundamental problems in statistical 
mechanics. It is also a very useful tool to tackle problems
in several fields of theoretical physics such as the study of
nonlinear dynamics and transitions to chaos \cite{hu},
disorder surface growth \cite{barabasi}, earthquakes \cite{didier},
among others. The conceptual foundation of the method, first laid by
Kadanoff \cite{kadanoff} to qualitatively predict scaling
behavior at a second-order phase transition, is to reduce,
in a step-by-step way, the degrees of freedom of the system 
leaving unchanged the underlying physics of the problem.
This reduction, carried out repeatedly through a renormalization 
recursion relation, leads the original
system with a large correlation length (the range at which
physical microscopic operators are correlated) to one with
correlation length of unity order, where well-known methods as
perturbation theory can, at least in principle, be used to
treat the problem. Depending on the mathematical technique
such thinning degrees of freedom can be implemented 
directly in the reciprocal (momentum) space or in the real (position)
space. The former approach makes use of mathematical tools
from quantum field theory with
the crystalline system being replaced by its continuous limit.
As a result, the so-called $\epsilon -$expansion proposed by
Wilson and Fisher \cite{wilson3} (and further developed by using
techniques of renormalized perturbation theory \cite{brezin,wallace})
provides analytical and quite well
controlled asymptotically exact results for critical exponents
(despite being unable in predicting values of critical
points, critical lines and phase diagrams). On the other hand, the more intuitive
real space version of the renormalization group works directly in
the position space. It was introduced by Niemeijer and
van Leeuwen \cite{leeuwen} and several different techniques have been proposed
and applied to a great variety of statistical models \cite{burkhardt}.
The real space renormalization group (RSRG) has since become an
important apparatus in studying critical phenomena. 

The main feature of the renormalization group
is to obtain, from the renormalization recursion relations,
flow diagrams in the parameter space
of the system, i.e., the space spanned by the different kinds of
interactions present in the Hamiltonian. The attractors (trivial fixed points) 
and their basins in the flow diagram correspond to regions of different 
thermodynamic phases. These regions are separated by critical frontiers 
associated to semi-stable attractors (relevant fixed points) which
determine the universality class of its critical exponents. More unstable
attractors located on the critical surfaces lead to multicritical frontiers 
(or, as it happens to be more usual, just multicritical points). There can also be a 
first-order frontier line linking a multicritical point to a first-order
fixed point located at infinite values of the Hamiltonian parameters. 
Moreover, the singularities of
the critical frontiers on the multicritical regions are characterized 
by the crossover exponents.
So, from the above general view, the topology of  flux diagrams 
allows one to achieve qualitative information about the critical
behavior of the system, e.g., universality,
order of transitions, crossover and multicriticality, among others.
Regarding the real space renormalization group approach, which will be the
subject of this work, one can also say that in most cases rather accurate 
quantitative values for the exponents and transition lines are obtained.

Just for completeness we should say that, from a different and more modern 
point of view, the renormalization group is essentially
based on the fractal structure exhibited by the system at the critical
point. A fractal, named by Mandelbrot \cite{mandelbrot}, describes structures
consisting of parts of any size (i.e., in any scale) 
which are similar to the whole.  
Consequently, the renormalization group commonly establishes a recursion relation
transforming the parameter space of the Hamiltonian into itself,
leaving thus the underlying physics unchanged. Due to this 
continuous scale invariance
a fractional dimension  is ascribed to the fractal (and, as a consequence, to
the critical exponents, as we shall shortly see in this Introduction),
in contrast to the integer
dimension of the translationally invariant Euclidian space. There are, however, 
systems exhibiting discrete scale invariance (a weaker form of scale
invariance symmetry which is not continuous) where dimensions or exponents
are complex. Even in such cases the renormalization group formalism turns 
out to be quite useful \cite{didier}. 
The interested reader is referred to the recent review by 
Sornette for further details concerning discrete scale invariance and complex
dimensions \cite{didier2}.

The renormalization group procedure has been extensively reviewed in
literature (see, for instance, the excellent reviews in References
\cite{wilson2,hu,wallace,burkhardt};
see also a more recent text by Yeomans on this
subject \cite{yeomans}). Regarding the  RSRG several techniques have been 
proposed and applied with success to various problems. Among the simplest 
ones we have decimation \cite{maris}, Migdal-Kadanoff \cite{m-k} and  
Niemeijer-van Leeuwen cells \cite{leeuwen} approaches, which have also
been discussed in the reviews referenced above. More accurate procedures
include Monte Carlo renormalization group \cite{swendsen0}, 
correlation function preserving renormalization group \cite{tsallis0}
and phenomenological renormalization group \cite{nightingale}.
It is not the scope of the present work to discuss all the methods above,
rather just the phenomenological ones will be treated. 
Nevertheless not less excellent reviews on
Monte Carlo renormalization group \cite{swendsen} and correlation function
phenomenological renormalization group \cite{tsallis} are addressed to the reader.

Despite its success, the ordinary RSRG has some drawbacks in the core of
its implementation. Although exact results are in general 
achieved in classical models
for dimension $d=1$, in most cases for $d>1$ some uncontrollable, and rather 
obscure, approximations have to be done in order to properly obtain
the renormalization recursion relations. In addition, one has also
to deal with the awkward ``proliferation of parameters", usually
encountered in approaches such as decimation \cite{maris}.

One class of  RSRG methods, free from the drawbacks presented above,
is the so-called ``phenomenological renormalization" or finite size
scaling approach. In these methods one computes (exactly) thermodynamic
quantities $P_L$ and $P_{L ^\prime}$ for two different finite systems 
and, from the expected scaling relation obeyed for this quantity $P$
in the limit where $L^\prime (< L)$ tends to infinity,
renormalization group recursion equations can be  obtained. 
In this way the critical properties of the infinite system are obtained 
(approximately) from the knowledge of the corresponding  properties of its
 finite lattice counterpart. 

The finite systems used in these phenomenological renormalizations depend 
strongly on the geometry of the lattice. For hypercubic lattices 
there are two specific geometries of particular interest:

i) a finite system in all directions consisting of a hypercube
of side $L$ in a  $d$-dimensional lattice; 

ii) a system infinite
in one direction and finite in the other $d-1$ dimensions with 
cross-sectional area characterized by the length $L$. 

Other geometries,
such as  systems infinite in two dimensions, can also be considered.
However, due to the difficulty in obtaining exact solutions even
for two-dimensional models, such geometry will not be discussed
here. On the other hand, boundary conditions applied to the finite 
systems depend on the particular renormalization group strategy. The 
common ones are the periodic boundary conditions and, in this case, just
bulk criticality is studied. There are, however, some methods which
allow to study bulk and surface criticality by taking free boundary
conditions in the finite lattices. Sometimes helical boundary conditions
are imposed in geometries ii) in order to obtain a sparse transfer
submatrix leaving the problem numerically more tractable \cite{candido}.

To illustrate, in a quite simple way, the procedure of writing
down a RSRG recursion relation for finite  size systems let us take,
without loss of generality, the magnetic language by means of the
spin-S Ising Hamiltonian model defined as
\begin{equation}
{\cal H }= -J \sum_{<i,j>} S_iS_j - h \sum_{i=1}^{\cal N} S_i~,
\label{eqn 0}
\end{equation}
where the first term is a sum  over nearest neighbors pairs of spins $<i,j>$
(which might be conveniently extended to more distant neighbors)
located on an arbitrary lattice and the second term is a single sum  over all 
the ${\cal N} \rightarrow \infty$ spins of the lattice. $J$ is the exchange interaction 
which is ferromagnetic for
$J>0$ or antiferromagnetic for $J<0$ and $h$ is an applied external magnetic 
field. The spin variables $S_i$ take values $S_i=S,S-1,...,-S$ 
where $S$ is a fixed integer or semi-integer positive number. 
This model exhibits a second-order phase transition at $K=K_c$ and $H=0$,
where $K=\beta J$ is  the reduced exchange interaction, $H=\beta h$ is the reduced
external magnetic field, $\beta = 1/k_BT$ and $k_B$ is the Boltzmann constant.
At zero external field  the correlation length $\xi $ close to the 
transition is expected to diverge as
\begin{equation}
\xi \approx |\epsilon| ^{-\nu}~,
\label{eqn 0a}
\end{equation}
where $\epsilon = K - K_c \approx 0$ and $\nu$ is the corresponding
critical exponent. 

From the Kadanoff block construction \cite{kadanoff}
it is shown that close to the transition the singular part of the free
energy per site $f_s(\epsilon,H)$ is a homogeneous function of its 
thermodynamic variables (as stated by Widom in 1965 \cite{widom}) i.e.,
\begin{equation}
f_s(\epsilon ,H) = \ell ^{-d}f_s(\ell ^{1/\nu } \epsilon ,
                  \ell ^{y_H}H)~,
\label{eqn 0aa}
\end{equation}
where $\ell$ is an arbitrary scaling factor, 
$d$ is the dimensionality of the corresponding lattice, and $y_H$ is
the magnetic critical exponent related to the magnetization $m$
through $m=H^{1/\delta}$, with $\delta = y_H/(1-y_H)$.
Now, any other quantity $P$, obtained as derivatives of the above 
free energy, will behave as power laws
\begin{equation}
P \approx |\epsilon| ^{-\sigma}~,
\label{eqn 0b}
\end{equation}
where $\sigma$ is the critical exponent of the quantity $P$. 
For example, the magnetization
\begin{equation}
m(K,H)={{Tr\left( {1\over {\cal N}}\sum_{j=1}^{\cal N}S_j\right)
\exp (-\beta {\cal H})}\over{Tr\exp (-\beta {\cal H})}}~,
\label{eqn 00b}
\end{equation} is obtained, close to criticality, from 
\begin{equation}
m(\epsilon ,H)=-{\partial f_s(\epsilon ,H) \over {\partial H}}=
\ell ^{-d+y_H}m(\ell ^{1/\nu } \epsilon ,\ell ^{y_H}H)~.
\label{eqn 0bb}
\end{equation}
At $H=0$ and choosing $\ell ^{-1/\nu} = \epsilon$ ($\epsilon >0$)
Eq. (\ref{eqn 0bb})
gives, after comparing to Eq. (\ref{eqn 0b}), 
$\sigma = \beta = -(d-y_H)\nu $.
Similarly, for the zero field specific heat 
$C=-T\left ( \partial ^2f_s/\partial T^2\right )_{H=0}$ and zero field
magnetic susceptibility 
$\chi = \left (\partial m/\partial H\right)_{H=0}$  it is easy to obtain 
$\sigma = \alpha = 2-d\nu$, 
and $\sigma =\gamma =(2y_H-d)\nu $, 
respectively. These power laws 
are a signature of the continuous scale invariance of the system 
at criticality.

On the other hand, a {\it finite} system of linear dimension 
$L$ close to $K_c$ will of course
present a non-diverging characteristic length and its
correlation length will, at the maximum, be given by $\xi \approx L$. Inserting
this relation in Eq. (\ref{eqn 0a}) and substituting for $\epsilon$
in Eq. (\ref{eqn 0b}) one gets
\begin{equation}
P \approx L ^{\sigma /\nu}=L^\phi~,
\label{eqn 0c}
\end{equation}
where $\phi =\sigma /\nu$ is called the anomalous dimension of the
quantity $P$. As $\phi$ is commonly a fractional number it is clear again
from the above relation the effects of continuous scale invariance on 
dimensions and exponents of critical systems.

According to the finite size scaling hypothesis, the generalized scaling  
relation obeyed by any thermodynamic quantity $P$ taking into account
the finite size $L$ of the system can be expressed as \cite{fisher,barber}
\begin{equation}
P(\epsilon  , H , L)=\ell ^{\phi }P(\ell ^{1/\nu } \epsilon ,\ell ^{y_H}H,
\ell ^{-1}L)~,
\label{eqn 1}
\end{equation}
where  $L\rightarrow \infty$ is the
linear dimension of the finite system and $\epsilon \approx 0, H \approx 0$.
By desiring the right hand side of Eq. (\ref{eqn 1}) to be the respective
quantity $P$ for a smaller system of linear dimension $L^\prime$ we choose
for the scaling factor $\ell ^{-1}L=L^\prime$ or
\begin{equation}
\ell={L\over L^\prime}~.
\label{eqn 2}
\end{equation}
Expressing now the variables 
$\ell ^{1/\nu } \epsilon =\epsilon ^\prime=K^\prime-K_c$ 
and $\ell ^{y_H}H=H^\prime$,
Eq. (\ref{eqn 1}) can be rewritten as
\begin{equation}
{P_{L^\prime }(K^\prime ,H^\prime ) \over L^{\prime \phi }}=
{P_L(K,H) \over L^\phi}~~,
 \label{eqn 3}
\end{equation}
where the system size is now given as a subscript instead of a variable
(to be more apparent) and
primed quantities refer to the smaller system  $L^\prime$.
This is the basic equation from which a mapping ($K,H$) $\rightarrow$
 ($K^\prime,H^\prime$) is obtained with the rescaling factor defined
in Eq. (\ref{eqn 2}). It is clear that the only source of inaccuracy
in this relation resides in the finite values taken for $L$ and $L^\prime$
in order to get  exact calculations for $P_L$ and $P_{L^\prime}$.
Nevertheless, it has been noted that
such approaches {\it are applicable to rather small systems of no more 
than a few degrees of freedom}.
Of course, the bigger the finite systems, the better the results achieved,
and it turns out that  sometimes they exceed in accuracy
more conventional methods such
as series expansion, $\epsilon$-expansion or other  renormalization
group procedures.

Different quantities $P$ will of course generate different approaches.
The first  procedure, and in our opinion one with the most
accurate results, was proposed by Nightingale \cite{nightingale} 
by using the correlation length computed for finite lattices 
consisting of infinite strips with finite width, i.e.,
the geometry  ii) discussed above. From the basic ideas of the renormalization 
group theory it is well known that the transformed system with less
degrees of freedom has a smaller correlation length which is the old one
reduced by the rescaling factor $\ell$, namely
\begin{equation}
\xi^\prime (K^\prime,H^\prime)={\xi(K,H) \over \ell}~.
\label{eqn 4}
\end{equation}
From equations (\ref{eqn 2}) and (\ref{eqn 4}) we then get
\begin{equation}
{\xi^\prime (K^\prime,H^\prime)\over L^\prime} = {\xi (K,H)\over L}~.
\label{eqn 5}
\end{equation}
This means that for the correlation length the anomalous exponent $\phi$
is equal to unity and the  above equation was interpreted by Nightingale
as a renormalization group transformation of the infinite system
from which fixed points, critical exponents, etc., are obtained.  
Since  Eq. (\ref{eqn 5}) involves just a single function, we obtain
just one scalar recursion relation instead of the usual multidimensional
recursion relation on a set of coupling constants such as $(K,H)$ in
the present example. So, flow diagrams analyses are not possible when the
Hamiltonian has more than two parameters. However, 
second-order transition lines
and estimates of critical exponents can be obtained by determining
a mapping 
\begin{equation}
K\rightarrow K^\prime =R(K;r)
\label{map1}
\end{equation}
for a fixed value of $r=H/K$, where the recursion relation $R(K;r)$ is,
in principle, obtained from Eq. (\ref{eqn 5}). The critical line is given
by the fixed point solutions $K^\prime = K = K^*=R(K^*;r)$ as a function of
$r$ and estimates of the critical  exponent is obtained from the 
linearization of $R$ around $K^*$
\begin{equation}
K^\prime - K^* = \lambda _T (K-K^*)~,
\label{map2}
\end{equation}
where the thermal eigenvalue $\lambda _T$ is given by
\begin{equation}
\lambda _T = \left. {{\partial K^\prime } \over {\partial K}}\right| _{K^*}=
\left. {{\partial R(K;r) } \over {\partial K}}\right|_{K^*}~,
\label{map3}
\end{equation}
with the corresponding thermal critical exponent
\begin{equation}
\ell ^{1/\nu }= \lambda _T~.
\label{map4}
\end{equation}
Estimates of the magnetic critical exponent are extracted from
\begin{equation}
\ell ^{y_H }= \left. {{\partial H^\prime } \over {\partial H}}\right|_{K^*}~,
\label{map5}
\end{equation}
from which, using Eq.(\ref{eqn 5}), gives
\begin{equation}
\ell ^{2y_H+1} = \left ( {{ \partial ^2 \xi ^\prime / \partial {H^\prime}^2}\over
{( \partial ^2 \xi / \partial {H}^2}}\right)_{K^*}~.
\label{map6}
\end{equation}
The second derivatives appear in obtaining $y_H$ since $\xi $ is commonly
an even function of $H$. The procedure outlined above is quite general when
using only one recursion relation and should be applied to any function
$R(K;r)$ (coming, for example,  from other renormalization group scheme) as 
well as to any couplings (not necessarily an external field).

Other thermodynamic functions (such as, for instance,  specific heat or 
magnetic susceptibility) can also be used to study the critical
properties of statistical mechanical systems through Eq. (\ref{eqn 3}). 
The main  problem  in these cases is that
the exponent $\phi$ is in general not known. The method can,
nevertheless, be implemented by force through the use of three different finite
systems ($L,~L^\prime,~L^{\prime\prime}$) and taking the
estimate of $\phi$ the value that yields the same fixed point
solution for the recursion relation (\ref{eqn 3}) from 
($L,~L^\prime$) and ($L^\prime,~L^{\prime\prime}$)
clusters, respectively \cite{santos}. In the present work we will
be concerned only with methods where $\phi$ has a known value.

By taking $P$ as the
order parameter of the system and together with mean field calculations
one can obtain the so-called mean field renormalization group (MFRG)
approach \cite{indekeu}.  Here, it is possible to overcome the
difficulty of not knowing the order parameter anomalous dimension $\phi$ and
finding out a recursion relation free from any exponent. 
This procedure will be discussed in more detail in the next section.
It is also possible, in some models  having more than one order parameter,
to obtain complete flux diagrams in the Hamiltonian coupling space
due to additional recursion relations. 
Moreover, by taking three clusters at the same time, it is allowed
to study bulk and surface critical behavior since the employed finite lattices 
must have open boundary conditions.

It is clear from Eq. (\ref{eqn 3})  that  in carrying out
phenomenological renormalization group calculations one desires to choose
quantities $P$ in which their anomalous dimension are known a priori.
A more recent proposal, for particular quantities having $\phi=0$,
has been introduced by de Oliveira in the finite size scaling renormalization
group (FSSRG) method \cite{oliveira,oliveiraa,neto}. This method is also capable to
circumvent, in some cases, the lacking of flow diagram lines in the 
parameter space by considering more than one quantity at the
same time and obtaining additional recursion relations, as in
some versions of the previous MFRG approach, even for models 
presenting just one order parameter. This approach will also be discussed
later on.

The purpose of this work is thus to review some phenomenological real space
renormalization group procedures namely, the mean field renormalization
group, the effective field renormalization group (EFRG) and the finite size
scaling renormalization group. We will not address here further details
on the Nightingale phenomenological renormalization since it has already
been widely discussed in literature (see, for instance, Reference
\cite{privman} and references therein). Special emphasis  will then be given
to the MFRG since it has been extended and improved in different ways and
applied to a vast class of models (classical, quantum, geometrical,
pure and diluted, static and dynamic). We will also not discuss here 
extensively the results achieved in each of its applications.
We will be mainly concerned here in all the versions of the method 
and its general formulations, some of them not previously published 
in the literature. Just a simple list of references in alphabetical order 
(without any comments) on works involving the use of MFRG up to 
the year  1993 has already been given by Croes and Indekeu  \cite{croes}.

The review is arranged as follows. In the next section we present
the formalism of the mean field renormalization group for the study
in static and dynamic critical phenomena. In Section III the exact results
from the MFRG are presented. Section IV gives an up to date application
of the method in static problems, and Section V does the same concerning
the dynamic critical behavior. Related phenomenological renormalization
group approaches
are presented in section VI and some final remarks are discussed in
section VII.

\section{Phenomenological Mean field renormalization group approach}

The phenomenological mean field renormalization group approach uses the
order parameter as the quantity to be renormalized. In 
magnetic models it is given by the magnetization of the system which
can generally be written as $m({\bf K},H)$, where the vector ${\bf K}$ 
represents all the reduced coupling constant interactions of the
Hamiltonian and $H$ is the reduced external magnetic field. It means that
the first sum in Eq. ({\ref{eqn 0}) may now not be restricted to
first neighbors but may also include more distant interactions as well as
additional fields. For instance, when ${\bf K}$ is written as
${\bf K}=(K_1,K_2,K_3,K_4,...)$, $K_1$ should represent nearest neighbor
interaction, $K_2$ next nearest neighbor interaction, $K_3$ a crystal
field interaction, $K_4$ a four spin interaction, and so on.
Since the MFRG has been proposed by Indekeu, Maritan and
Stella in 1982 \cite{indekeu}
the method has been widely used in treating several statistical mechanical
problems as well as frequently improved in different aspects of its formulation.
This section is devoted to review its basic assumptions and 
discuss some new implementations.

\subsection{MFRG}

In the original MFRG one  first considers two clusters of interacting spins
containing $N$ and $N^\prime$ sites, with $N^\prime<N$. The surrounding
spins of these clusters are fixed to a value $b$ and $b^\prime$, 
respectively, which, in a sense, can be viewed as effective
magnetizations representing the remaining spins of 
the infinite lattice and behave as a symmetry breaking field in each cluster.
The magnetizations per site $m_N({\bf K},H;b)$ and 
$m_{N^\prime}({\bf K^\prime},H^\prime;b^\prime)$ can exactly be computed
from
\begin{equation}
m_{N}({\bf K},H;b)={{Tr\left( {1\over N}\sum_{j=1}^NS_j\right)
\exp (-\beta {\cal H}_N)}\over{Tr\exp (-\beta {\cal H}_N)}}~,
\label{eqn 5a}
\end{equation}
where the trace is taken over the ensemble defined by the cluster Hamiltonian
${\cal H}_N({\bf K},H;b)$ and $S_j$ is the corresponding spin operator. 
Similar expression holds for the smaller cluster $N^\prime$.
The usual mean field type approximation is obtained by
assuming $m_N({\bf K},H;b)=b$ (or  
$m_{N^\prime}({\bf K^\prime},H^\prime;b^\prime)=b^\prime$) from which
$m=b$ (or $m^\prime=b^\prime$) is solved self-consistently at $H=0$ 
($H^\prime=0$) and the criticality is identified with the bifurcation
point in the above equations. Although this mean field approximation improves
the critical couplings as the cluster size gets larger, 
the critical exponents  are always the classical ones (in particular,
for the Ising model described in Eq. (\ref{eqn 0}) one has
$\alpha =0$, $\beta =1/2$ and $\gamma = 1$ for any lattice dimension).

Instead of the naive assumption of just equalling the symmetry breaking 
fields to the magnetization per spin for each cluster, the MFRG assumes
that these magnetizations are related through the scaling relation given by
Eq. (\ref{eqn 0bb}) for the order parameter of the infinite system 
close to criticality \cite{leeuwen}
\begin{equation}
m_{N^\prime}({\bf K^\prime},H^\prime;b^\prime)=
\ell^{d-y_H}m_{N}({\bf K},H;b)~,
\label{eqn 6}
\end{equation}
where the rescaling factor $\ell$ can be defined by the usual way
\begin{equation}
\ell={(N/N^\prime)}^{1/d}~,
\label{eqn 7}
\end{equation}
which is equivalent to definition (\ref{eqn 2}) for symmetric blocks
of size $N=L^d$.
As Eq. (\ref{eqn 6}) should hold for systems near
the critical point, the magnetizations for each finite
cluster must be very small. 
Such condition is achieved by letting $b\ll1$ and $b^\prime\ll1$, as well as
$H\ll1$ and $H^\prime\ll1$, since any finite system has no spontaneous
magnetization. Equation (\ref{eqn 6}) can thus be expanded and, to lowest order
in $b,~b^\prime$ and $H,~H^\prime$ gives
\begin{equation}
g_{N^\prime}({\bf K^\prime})H^\prime +
f_{N^\prime}({\bf K^\prime})b^\prime=\ell^{d-y_H}f_N({\bf K})b
+\ell^{d-y_H}g_N({\bf K})H~,
\label{eqn 8}
\end{equation}
where
\begin{equation}
f_{N^\prime}({\bf K^\prime})=\left.
{{\partial m_{N^\prime}({\bf K^\prime},H^\prime;b^\prime)}\over
{\partial b^\prime}}\right|_{H^\prime=0,b^\prime=0}  
~~~;~~~~~~~~
f_N({\bf K})=\left.
{{\partial m_{N}({\bf K},H;b)}\over
{\partial b}}\right|_{H=0,b=0}~,
\label{eqn 9}
\end{equation}
\begin{equation}
g_{N^\prime}({\bf K^\prime})=\left.
{{\partial m_{N^\prime}({\bf K^\prime},H^\prime;b^\prime)}\over
{\partial H^\prime}}\right|_{H^\prime=0,b^\prime=0}  
~~~;~~~~~~~~
g_N({\bf K})=\left.
{{\partial m_{N}({\bf K},H;b)}\over
{\partial H}}\right|_{H=0,b=0}~.
\label{eqn 9i}
\end{equation}

As $b$ and $b^\prime$ are also viewed as magnetizations and are
very small they are assumed to satisfy the same  scaling relation
given by Eq. (\ref{eqn 6}), i.e.,
\begin{equation}
b^\prime=\ell^{d-y_H}b~.
\label{eqn 10}
\end{equation}
From equations (\ref{eqn 8}) and (\ref{eqn 10}) one then gets
\begin{equation}
f_{N^\prime}({\bf K^\prime})=f_N({\bf K})~,
\label{eqn 11}
\end{equation}
obtaining by equalling the coefficients of $b^\prime $ and
\begin{equation}
g_{N^\prime}({\bf K^\prime})H^\prime=\ell^{d-y_H}g_N({\bf K})H~,
\label{eqn 11i}
\end{equation}
obtaining by equalling  the external field terms.
Equation (\ref{eqn 11}), which is independent of any exponent, is interpreted
as a recursion relation for the coupling constants {\bf K} according
to the ordinary MFRG, while equation(\ref{eqn 11i}), as  shown
below, is used to estimate the magnetic critical exponent $y_H$
(the recursion relation does not depend on the external field). 
Clearly, for a multidimensional Hamiltonian 
parameter space this scalar recursion relation (\ref{eqn 11})
does not provide complete 
flux diagrams. This sole equation can, however, be used to
estimate critical exponents and locate critical surfaces by considering 
a mapping $K_1\rightarrow K_1^\prime =R(K_1,r_1,r_2,...)$ for fixed 
values of the ratio $r_1=K_2/K_1$, $r_2=K_3/K_1$, ..., where 
${\bf K}=(K_1,K_2, ...)$ and the function R is, in principle,
obtained from the rescaling relation (\ref{eqn 11}). By computing the fixed
point solution $K_1=K_1^\prime=K_1^*=R(K_1^*,r_1,r_2, ...)$ as a
function  of $r$ one determines 
the critical surface, represented here as ${\bf K}^*=(K_1^*,r_1, r_2,...)$, 
and the corresponding exponents are achieved through
\begin{equation}
\nu ={{\ln \ell}\over {\ln \lambda_T}}~,
\label{eqn 12}
\end{equation}
as stated by Eq. (\ref{map4}}) for the correlation length critical exponent where
\begin{equation}
\lambda_T=\left.
{{\partial K_1^\prime}\over {\partial K_1}}\right|_{{\bf K}^*}=\left.\left(
{{\partial f_N}\over \partial K_1} {\left({\partial f_{N^\prime}\over\partial
K_1^\prime}\right)}^{-1}\right)\right|_{{\bf K}^*}~,
\label{eqn 13}
\end{equation}
and
\begin{equation}
g_{N^\prime}({\bf K}^*)=\ell^{d-2y_H}g_N({\bf K}^*)~
\label{eqn 14}
\end{equation}
for the magnetic critical exponent $y_H$, where the relation
 $H^\prime=\ell^{y_H}H$ has been used in Eq. (\ref{eqn 11i}).

This approach has been applied to a great variety of statistical
models and quite good results have been obtained even by taking
the simplest choice for the clusters, namely, $N^\prime=1$ and
$N=2$.  Some exact results are also obtained from this method when
treating some specific systems. A summarized discussion of such
studies will be given in the next sections.
However, in order to taste the simplicity of the method and its
potentialities let us apply it to three different problems by
taking the smallest clusters depicted in Fig. 1.

{\it spin-1/2 nearest neighbors Ising Model}

The Hamiltonians ${\cal H}_1$ and ${\cal H}_2$ for the clusters of one
and two spins shown in Figs. 1(a) and 1(b) can be written as
\begin{equation}
{\cal H}_1= -cJ^\prime b^\prime S_1 - h^\prime S_1~,
\label{eqn h1}
\end{equation}
\begin{equation}
{\cal H}_2=-JS_1S_2-(c-1)Jb(S_1+S_2)-h(S_1+S_2)~,
\label{eqn h2}
\end{equation}
where $S_i=\pm 1$ and $c$ is the coordination number of the lattice.
From the definition of the magnetization (\ref{eqn 5a}) one easily has
\begin{equation}
m_1=\tanh (cK^\prime b^\prime + H^\prime)~,
\label{eqn m1}
\end{equation}
\begin{equation}
m_2={{\sinh [2(c-1)Kb+2H]} \over {\cosh  [2(c-1)Kb+2H] + e^{-2K}}}~,
\label{eqn m2}
\end{equation}
which, for small values of $b,b^\prime ,H $ and $H^\prime $ reduce to
\begin{equation}
m_1=cK^\prime b^\prime + H^\prime~,
\label{eqn m1e}
\end{equation}
\begin{equation}
m_2={{2(c-1)Kb} \over {1 + e^{-2K}}}+ {{2H} \over {1 + e^{-2K}}}~.
\label{eqn m2e}
\end{equation}
The usual mean field approximation assumes $m_1=b^\prime $ and $m_2=b$
yielding, at zero external field, $K_c=1/c$ and $1 + e^{-2K_c}=2(c-1)K_c$.
One can note that from both clusters there is a (wrong) finite critical 
temperature $T_c\ne 0$ for the one-dimensional model ($c=2$). 
For a future comparison to the two-dimensional model ($c=4$) 
the later expressions give $K_c=0.250$
and $K_c=0.286$ for $N^\prime =1$ and $N=2$, respectively.
Moreover, by further expanding equations (\ref{eqn m1}) and (\ref{eqn m2})
up to order ${b^\prime}^3$ and $b^3$ it is easy to show that the
corresponding critical exponents take the classical values,
independent of the dimensionality of the lattice. 

Reminding now equations (\ref{eqn 8}-\ref{eqn 11i}) the corresponding MFRG
recursion relation
\begin{equation}
cK^\prime = {{2(c-1)K} \over {1+e^{-2K}}}
\label{eqn rrm}
\end{equation}
gives
\begin{equation}
K_c={1\over 2}\ln {1 \over {c-2}}~,
\label{eqn Kc}
\end{equation}
where $K_c=K^\prime =K$ is the non-trivial fixed point solution of the equation
(\ref{eqn rrm}). The above result is the same as that obtained from the 
Bethe approximation \cite{bethe}. The critical exponents according to equations 
(\ref{eqn 13}) and (\ref{eqn 14}) are
\begin{equation}
\ell ^{1/\nu }= 1 + {{(c-2)} \over {2(c-1)}}\ln {c\over {c-2}}~,~~~~~~
  \ell ^{d-2y_H}={(c-1) \over c}~,~~~~~\ell= 2^{1/d}~.
\label{eqn exp}
\end{equation}
In this case, for the one-dimensional model ($c=2$) one has the exact critical 
temperature $T_c=0$ and critical exponent $y_H=1$
(these values are in fact obtained for any cluster
sizes), while the critical exponent $\nu =1$ is only achieved by comparing two
chains with $N$ and $N-1$ spins in the limit $N \rightarrow \infty $.
Numerical values for the two-dimensional model are $K_c=0.347$,
$\nu=1.667$ and $y_H=1.415$, which are, when compared to the
previous mean field results (and without further effort),
closer to the exact ones, mainly for the critical coupling
(see Table \ref{tab 1}). The corresponding results for the $d=3$ lattice 
using the present clusters are listed in Table \ref{tab 2}.

{\it Site percolation }

The site percolation \cite{stauffer0} is a geometrical problem
defined on a regular infinite lattice 
where each site can independently be occupied or empty with probabilities $p$
and $1-p$, respectively ($p$ is also viewed as the concentration
of the occupied sites). A  {\it cluster}  is formed by
grouping together nearest neighbors occupied sites. At $p=1$ one has
an infinite cluster linking all sites. On the other hand, for small
$p$ one has an infinite number of independent finite clusters. Then there is a 
critical concentration $p_c$ above which the size of the clusters 
(at least one) becomes
infinite. The mean linear size of the clusters is the correlation
length $\xi$ which, close to the critical threshold diverges as
\begin{equation}
\xi \approx |p-p_c|^{-\nu _p}~,
\label{eqn p1}
\end{equation}
where $\nu _p$ is the percolation correlation length critical exponent.
$p_c$ depends on the topology of the lattice and $\nu _p$ depends only
on the lattice dimensionality.

Let us now consider the same clusters (a) and (b) in Fig. 1 for this problem
\cite{bell1,bell2}. The order parameter $P_N$ is defined by the probability 
of a particular site to belong to the infinite cluster. $b$ and $b^\prime$
will be now the effective values of the probability, for surrounding sites 
of each block, to belong to the infinite cluster. For the one
site cluster ($N^\prime =1$) its probability of belonging to the infinite
cluster is $p^\prime b^\prime$ times the number $c$ of surrounding sites, i.e.
\begin{equation}
P_1 = cp^\prime b^\prime~.
\label{eqn pb1}
\end{equation}
In a similar way, the probability of any of the two sites of the $N=2$ cluster 
is, up to first-order in $b$, given by
\begin{equation}
P_2=p(1-p)(c-1)b+2p^2(c-1)b=p(1+p)(c-1)b~.
\label{eqn pb2}
\end{equation}
The first term on the left hand side of the above equation 
takes into account the contribution
of the configuration  where one site of the cluster is occupied and
the other one is empty, and the second term the contribution
when both sites are occupied. In all terms just configurations with
one surrounding site occupied with probability $b$ are taken since
configurations presenting two or more mean field sites will contribute
to the order parameter a factor of order $b^2$ or greater.
Mean field approximation ($P_1=b^\prime $ and $P_2 =b$) gives
$p_c=1/c$ for the smaller cluster and 
$ p_c={1 \over 2} \left( { \sqrt{{c+3}\over {c-1}} -1}\right)  $ 
for the bigger one, while the MFRG furnishes the recursion relation
\begin{equation}
p^\prime c=p(1+p)(c-1)~,
\label{eqn rrp}
\end{equation}
from which one obtains the non trivial fixed point $p=p^\prime =p_c$
\begin{equation}
p_c={1\over {c-1}}~,
\label{eqn pf}
\end{equation}
again identical to the Bethe approximation \cite{sarbach}. The
corresponding critical exponent is given by
\begin{equation}
\ell ^{1/\nu _p}= {{c+1}\over c}~.
\label{eqn ce}
\end{equation}
As in the previous case, for the one-dimensional model the mean field
approximation results $p_c <1$, contrary to the exact value $p_c=1$.
This exact result is however obtained from the MFRG recursion
relation (\ref{eqn pf}). The correct value of $\nu_p=1$ is achieved
only in the limit of infinite chains. For the two-dimensional 
square lattice equations (\ref{eqn pf}) and (\ref{eqn ce}) furnish
$p_c=1/3$ and $\nu=1.55$, which should be compared to the series
values $p_c=0.59$ and $\nu=1.33$ \cite{essam}. The results
are analogous for $d=3$. In addition,
the related problem of bond percolation where bonds  are independently
present linking  two neighboring sites with probability $p$ or absent
with probability $1-p$ can similarly be treated from this formalism.
It turns out that the results are identical to those above for the
site problem when using the smallest clusters. Only by taking bigger
finite systems the approach is able to distinguish between the site percolation
and the bond percolation problems.

{\it Directed self-avoiding random walk}

Let us briefly mention the directed self-avoiding random walk since 
it is, in some sense, related to the site percolation problem treated above.
In order to define the directed self-avoiding random walk (DSARW) let us 
consider a $d$-dimensional hypercubic lattice \cite{manna}. The DSARW
is the geometrical object built by starting at a given point (the origin)
and advancing at random through steps linking nearest neighbors sites,
with the constraint that any step be only along the positive directions of
the cartesian axis. Letting $w$ be the fugacity associated with a step,
one has that for $w$ larger than a critical value $w_c$ the walk will, on
average, extend to an arbitrarily large distance.
In this case one can distinguish two different directions: one parallel 
($\parallel$) defined along the bisectrix of the  angle formed by the cartesian 
axis, and another perpendicular ($\perp$) taken orthogonally to the previous one.
It turns out that as $w$ approaches $w_c$ the root mean square of the 
parallel and perpendicular displacements diverge as power laws of the form
$(w-w_c)^{\nu_\parallel}$ and $(w-w_c)^{\nu_\perp}$, respectively, with 
$\nu_\parallel$ and  $\nu_\perp$ being the corresponding critical indices.
Figs. 1(c) and 1(d) illustrates the smallest possible clusters which can
be used by the MFRG in order to treat the DSARW on a two-dimensional lattice. 

Defining now the order parameter $P_N$ as the generating function of all
walks that begin at the origin site and end at the mean field sites (with 
$b$ being the effective value of this order parameter for the surrounding
sites)  one arrives at 
\begin{equation}
P_1=2w^\prime b^\prime~,
\label{eqn dp1}
\end{equation}
since there are two ways with probability $w^\prime b^\prime$ of linking 
the origin of this smallest cluster to its two boundary sites and
\begin{equation}
P_2=4w^2b~,
\label{eqn dp2}
\end{equation}
where for this cluster one has four different ways of linking the origin to the
three boundary sites all having the same probability $w^2b$.
The MFRG recursion relation is then $w^\prime = 2w^2$
from which one has the exact non trivial fixed point $w_c=1/2$.
Exact values for the critical indexes (which are more laborious to be
computed) $\nu_\parallel =1$ and
$\nu_\perp =1/2$ are also obtained (in fact, for any cluster size, as well
as for any dimension $d$. See Ref. \cite{neves} for more details).

Although in all models above the calculation of the order 
parameters could be readily performed for the smallest clusters, larger
systems will drastically increase the analytical and computational effort. 
Tables \ref{tab 1} and \ref{tab 2}
contain the results obtained for the spin-1/2 Ising model with
nearest neighbor interactions up to the greatest size treated so far 
($L=6$ for the two-dimensional model and $L=3$ for the three-dimensional one),
and Table \ref{tab 2a} the corresponding values for the two-dimensional
site percolation problem.
A more detailed discussion of the values in these tables  will be 
given in the next subsection.

Despite its simplicity the usual MFRG treated above deserves some critics
regarding the use of larger clusters. Since the symmetry breaking fields $b$ and
$b^\prime$, in some sense, take into account the remaining of the
infinite lattice, the scaling imposed for the magnetizations
$m_{N^\prime}$ and $m_N$ is the relation obeyed by the {\it infinite} system 
order parameter  (note that equation
(\ref{eqn 6}) comes from the Wilson renormalization group strategy). 
On the other hand, the
final expression of the MFRG, namely Eq. (\ref{eqn 11}), is
independent of the symmetry breaking fields. From this equation the functions 
$f_N$ and $f_{N^\prime}$ can be thought of being quantities satisfying
the finite size scaling relation (\ref{eqn 3}) with anomalous dimension
$\phi =0$. However, for very large clusters such relation for
$f_N$ and $f_{N^\prime}$ (with $\phi=0$) is not true. 
This is most easily seen by taking 
the limit of $N$ and $N^\prime$ going to infinity. It is clear that 
while the computed magnetizations are bulk quantities, the surrounding 
effective magnetizations on each cluster behave as surface quantities 
and must  properly scale as surface
fields. In this limit Eq. (\ref{eqn 10}) is no longer valid
once $y_H$ is the bulk magnetic critical exponent.
As a consequence, the procedure will not reproduce, in general, 
the exact results  as the  size of the systems tend to infinity.

\subsection{Surface-Bulk MFRG (SBMFRG)}

A way to handle this situation has also been proposed by Indekeu
and co-workers \cite{indekeu2}. For large clusters, except for spins
at corners or edges, almost all boundary spins are subject to the
same effective field which is proportional to the surrounding
fixed magnetizations. According to a rigorous finite
size scaling derivation these surrounding fields do not scale like
a magnetization itself but like a surface field \cite{binder}.
So, for $b$ and $b^\prime$ one has 
\begin{equation}
b^\prime=\ell^{y_{HS}}b~,
\label{eqn 15}
\end{equation}
where $y_{HS}$ is the corresponding surface field critical exponent.
One can note that now the factor $\ell ^{d-y_H}$ is no more 
eliminated from equations (\ref{eqn 8}) and (\ref{eqn 15})
and a simple formulation, as expressed by Eq. (\ref{eqn 11}),
is not available. It is
needed then three clusters $N$, $N^\prime$ and $N^{\prime \prime}$
(with $N>N^\prime > N^{\prime \prime}$),
instead of two, in order to get a renormalization group transformation.
In this case, from $m_N({\bf K},H;b)$, 
$m_{N^\prime}({\bf K^\prime},H^\prime;b^\prime)$ and
$m_{N^{\prime\prime}}({\bf K^{\prime\prime}},H^{\prime\prime};
b^{\prime\prime})$ one computes $f_N({\bf K})$,
$f_{N^\prime}({\bf K^\prime})$, 
$f_{N^{\prime\prime}}({\bf K^{\prime\prime}})$ and from Eq.
(\ref{eqn 8}) at $H=H^\prime =H^{\prime \prime }=0$ together with
\begin{equation} 
b^\prime=\ell_{NN^\prime}^{y_{HS}}b~~,~~~~~~~~~~~~ 
b^{\prime\prime}=\ell_{N^{\prime}N^{\prime\prime}}^{y_{HS}}b^\prime ~,
\label{eqn 16}
\end{equation}
it is easy to obtain
\begin{equation}
f_{N^\prime}({\bf K^\prime})=\ell_{NN^\prime}^{d-y_H-y_{HS}}f_N({\bf K})~,
\label{eqn 16a}
\end{equation}
\begin{equation}
f_{N^{\prime\prime}}({\bf K^{\prime\prime}})=
\ell_{N^{\prime}N^{\prime\prime}}^{d-y_H-y_{HS}}f_{N^\prime}({\bf K^\prime})~,
\label{eqn 16b}
\end{equation}
where
\begin{equation}
\ell_{NN^\prime}=(N/N^\prime)^{1/d}~,~~
\ell_{N^{\prime}N^{\prime\prime}}=(N^\prime/N^{\prime\prime})^{1/d}~.
\label{eqn 17}
\end{equation} 
Following the well-established optimization strategy \cite{santos}
the exponent $d-y_H-y_{HS}$ is determined self-consistently by imposing
that the two different mappings with rescaling factors $\ell_{NN^\prime}$
and $\ell_{N^{\prime}N^{\prime\prime}}$ possess the same fixed point
${\bf K}={\bf K}^\prime={\bf K}^{\prime \prime}={\bf K}^*$.
This procedure allows one to obtain unambiguously the index $d-y_H-y_{HS}$
while two different estimates  for the critical exponents
$\nu$, $y_H$ and $y_{HS}$ are achieved from each of the renormalization
group transformations (\ref{eqn 16a}) and (\ref{eqn 16b}) (there are in
fact three different estimates since from equations (\ref{eqn 16a}) and 
(\ref{eqn 16b}) one can construct an additional relation between 
$f_{N^{\prime \prime}}$ and $f_{N}$). 

At this point it is worthwhile to see a quantitative and comparative
application of these methods in the well known control system square 
two-dimensional Ising model. Table \ref{tab 1} shows the critical temperature and
critical exponents for different growing finite size square clusters. It is
clear that a better accuracy is achieved from SBMFRG as compared to the
usual MFRG, mainly for larger clusters, where the approach to the exact
results is rather apparent. However, concerning small clusters,
the distinction of both methods is not so profound. First,
in assumption (\ref{eqn 11})  the ordinary MFRG assumes a zero value for
the critical index $d-y_H-y_{HS}$ (the exact one being $-0.375$).
From the SBMFRG one gets the best value $d-y_H-y_{HS}=-0.217$ 
for the biggest systems while
for the simplest choice $N=4,~N^\prime=2$ and $N^{\prime \prime }=1$ 
one has $-0.124$ \cite{indekeu2}, which is a  quite small  value. 
Second, for small clusters
there is no even a clear distinction between surface and bulk quantities.
This means that the fact of the SBMFRG being exact when 
$N$, $N^\prime$ and $N^{\prime \prime} \rightarrow \infty$ does not
make it more correct than the original MFRG for small systems.
Which method should be applied to a particular problem must be
decided on practical grounds. Table \ref{tab 2} lists the results for the $d=3$
Ising model and Table \ref{tab 2a} the values for the $d=2$ site percolation
problem.

Another  point deserving some discussion concerns the definition
of the rescaling factor $\ell$. As pointed out by Slotte \cite{slotte}
the rescaling factors given by equations (\ref{eqn 7}) and (\ref{eqn 17})
are, to some extent, arbitrary. Different definitions strongly affect the
results of the critical exponents for small clusters. 
Slotte suggested a definition of the rescaling factor where the
length of the cluster is measured by the number of interactions, including 
those with the surrounding mean field sites, instead of just
the effective number of sites, i.e.,
\begin{equation}
L=\left({1\over d}\sum_iL_i^{-2}\right)^{-{1\over 2}}~,
\label{eqn sf}
\end{equation}
where the sum runs over cartesian directions and $L_i$ are the number
of bonds along each direction. The one spin cluster has then $L=2$ and
the two-spin cluster $L=6\sqrt {d/(9d-5)}$. Such definition improves
substantially the critical exponents \cite{slotte}.
There are, however,
in the original MFRG, some quantities that are independent of $\ell$.
One of them is clearly the fixed point ${\bf K}^*$, as can be seen 
from Eq. (\ref{eqn 11}). The other one is the product
$\nu y_H$. By abandoning the definition given by Eq. (\ref{eqn 7})
the relation (\ref{eqn 14}) can be rewritten as 
 \begin{equation}
N^\prime g_{N^\prime}({\bf K}^*)=\ell^{-2y_H}Ng_{N}({\bf K}^*)~,
\label{eqn 18}
\end{equation}
which gives
\begin{equation}
y_H={1\over2}{{\ln Ng_N({\bf K}^*)/N^\prime g_{N^\prime}({\bf K}^*)}\over
{\ln \ell}}~,
\label{eqn 19}
\end{equation}
where $\ell ^d$ in (\ref{eqn 14}) has been replaced by $N/N^\prime$.
While the exponents $\nu$ and $y_H$ separately depend on the rescaling
factor, it is easy to see now from equations (\ref{eqn 12}) and 
(\ref{eqn 19}) that the product $\nu y_H$ is independent of $\ell$,
removing thus a principal weakness of the MFRG. In the
unified SBMFRG this is no longer  strictly true, but should still hold
to a first approximation. It is then expected that the critical exponent product
$\nu y_H$ will present a better precision than the exponents themselves.
Indeed, from the last column of Table I one can see that the best
product $\nu y_H$ from SBMFRG is  within $1.8 \%$ of error from its 
exact value, whereas the error in critical exponents ranges between 
$7.8\% - 9.4 \%$. The figures for MFRG are, respectively, $3\%$
and $14\%-16\%$. Roughly the same behavior is noticed in the data of Table
\ref{tab 2} for the $d=3$ Ising model.

Similar results are also obtained in an extended version of
the present approach by including surface and corner critical
exponents in two-dimensional models \cite{indekeu3}. In such
$d=2$ systems the distinction between surface and corner fields 
is easily made. All one has to do is to include an additional
corner field critical exponent $y_{HC}$ in the formalism above. Table \ref{tab 3}
quotes the results from Croes and Indekeu \cite{croes} for
the two-dimensional square Ising model. In this reference one also
finds critical values for other two-dimensional lattices namely,
triangular, honeycomb, hexagonal as well as the square lattice
with next nearest neighbors interactions. On the other hand, additional
edge fields will be present in three-dimensional finite 
lattices. However, bigger finite systems will require much
computational effort and inclusion of edge fields 
would not be so relevant in rather small systems.

\subsection{Extended MFRG (EMFRG)}

In the previous subsections just Hamiltonian models with one  parameter
have been discussed where the trivial one-dimensional renormalization
group flux is obtained. Let us now turn our attention to the case where
the Hamiltonian presents more than one parameter (or coupling constant).
It is clear that in this case  the
complete renormalization flow diagram is not defined through just the
one-dimensional recursion relation as given by Eq. (\ref{eqn 11}).  In the
particular two-dimensional case where we have ${\bf K}=(K_1,K_2)$,
which is the most studied in literature and will be treated in this section,  
estimates of critical lines and critical exponents for
models presenting only one order-parameter are often obtained through the
mapping $K_1\rightarrow K_1^\prime=R(K_1,r)$, for fixed values of the ratio
$r=K_2/K_1$, as has already been discussed in the Introduction.  
That is what one finds,
for instance, in the random Ising model or the transverse Ising model.  However,
depending on the system we are studying, other additional patterns may occur
where the above simple scheme cannot be applied for a two-fold coupling space
(in some models even for a one order-parameter system).
In these cases the thermodynamic quantities  should be considered in 
a proper way. This subsection is devoted to present a detailed discussion
of what has been done in overcoming such debility of the method as well as
to present an attempt to unify the approaches done so far by using the MFRG.
The formalism is intended to be quite general. However, whenever possible,
some reference to particular models (without presenting  
results; the corresponding reference will be listed in section IV)
exhibiting the behavior under consideration will be made just for clarity
and as a matter of example.  

\subsubsection{one order-parameter}

For systems like, for example, the antiferromagnetic spin-1/2
Ising model in an external field (recall Eq. (\ref{eqn 0}))
\begin{equation}
{\cal H }= J \sum_{<i,j>} S_iS_j - h \sum_i S_i~,~~~~~S_i=\pm1 ~,
\label{eqn afm}
\end{equation}
where $K_1=\beta J$ and $K_2=\beta h$
and the spin-1 Blume-Capel \cite{blume,capel} model
\begin{equation}
{\cal H }= -J \sum_{<i,j>} S_iS_j - \Delta \sum_i S_i^2~,~~~~~S_i=\pm1 ,0 ~,
\label{eqn bcm}
\end{equation}
where $K_1=\beta J$ and $K_2=\beta \Delta$ (the latter interaction being
the crystal field anisotropy), one has two couplings 
${\bf K}=(K_1,K_2)$ and just one
order-parameter, namely, the difference $m_+-m_-$ of the sublattice 
magnetizations $m_+$ and $m_-$ in the former case, and the mean 
value of the spin $<S_i>$ at any site  of the lattice in the
later model.  Thus, for a finite cluster of $N$ sites one can compute the
order-parameter $O_1=O_1({\bf K},b,q)$ and the other non-critical variable
$O_2=O_2({\bf K},b,q)$ (in the above examples $O_2$ is the net magnetization 
$m_++m_-$ of the antiferromagnet and the mean value of the square of the spin 
$<S_i^2>$ in the Blume-Capel model).  $b$ is the field conjugated to 
the order-parameter  $O_1$ and $q$ the
respective field conjugated to the non-critical variable $O_2$.  Close to the
transition just $O_1$ is small so the expressions above are expanded for small
$b$ to give 
\begin{equation} 
O_1=F_1({\bf K},q)b~, 
\label{eqn 20} 
\end{equation}
\begin{equation} 
O_2=O_2({\bf K},b=0,q)=F_2({\bf K},q)~.  
\label{eqn 21}
\end{equation} 
Note that as $b\rightarrow 0$ one has $O_1 \rightarrow 0$ and $O_2\ne 0$,
irrespective of the value of $q\ne 0$, whereas $O_2 \rightarrow 0$ when one further has
$q\rightarrow 0$.  Similarly, for a smaller cluster of $N^\prime$ sites one
finds 
\begin{equation} 
O_1^\prime=F_1^\prime({\bf K^\prime},q^\prime)b^\prime~,
\label{eqn 22} 
\end{equation} 
\begin{equation} 
O_2^\prime=O_2^\prime({\bf
K^\prime},b^\prime=0,q^\prime)= F_2^\prime({\bf K^\prime},q^\prime)~.
\label{eqn 23} 
\end{equation} 
From now on we will omit the subscripts $N$ and
$N^\prime$ and just denote the corresponding thermodynamic functions for
different clusters by unprimed and primed quantities, respectively.  According to
the ordinary MFRG one then finds 
\begin{equation} 
F^\prime_1({\bf K^\prime},q^\prime)=F_1({\bf K},q)~, 
\label{eqn 24} 
\end{equation} 
which, although independent of any exponent and rescaling factor, {\it depends 
on the additional fields $q$ and $q^\prime$}. It should be said that
the above behavior is not quite general. Counter examples are, among others,
disordered systems \cite{droz,pla0} and some quantum spin  models
\cite{pla0} where, despite having a two parameter Hamiltonian, they
present no non-critical variables  being unnecessary the
introduction of extra fields $q$ and $q^\prime$. As a result, Eq. 
(\ref{eqn 24}) is just dependent on ${\bf K}$ and ${\bf K^\prime}$
and the procedure $K_1\rightarrow K_1^\prime=R(K_1,r)$, for fixed values 
of the ratio $r=K_2/K_1$, outlined in the preamble of this subsection 
can easily be implemented. 

In studying the antiferromagnetic Ising
model in the triangular lattice Slotte \cite{slotte2} has proposed the mean
field Ansatz where $q$ and $q^\prime$ are self consistently obtained by
requiring that $O^\prime_2=q^\prime~$ and $~O_2=q$. Reasonable results, specially
for the value of $K_c$ have been achieved.  This approach has also been used in
the study of the Blume-Capel model \cite{osiel} where results, {\it not consistent
with the expected ones} (mainly by taking the smallest cluster sizes) have been
obtained, including  an unexpected phase transition at a finite critical 
temperature for the one-dimensional version (as seen in section II.A the
exact result is expected to be obtained from smallest clusters at least
for the one-dimensional model).  

A different choice, 
and equally natural, for the size dependence of the non-critical variable 
has been proposed by Plascak and S\'a Barreto \cite{pla1} and successfully 
applied to the Ashkin-Teller model. As this variable has its own size 
dependence which is not governed by finite size scaling at the transition 
it is assumed a {\it renormalization group} Ansatz in a way that
$q^\prime =q$, $O^\prime_2=O_2$, or
\begin{equation}  
F^\prime_2({\bf K^\prime},q)= F_2({\bf K},q)~,
\label{eqn 25}
\end{equation}
from which $q=q^\prime$ is obtained as a function
of ${\bf K}$ and ${\bf K^\prime}$ and inserted in Eq. (\ref{eqn 24}).
This renormalization group Ansatz has been proved to be better than the
mean field one for the antiferromagnetic Ising
model\cite{slotte3} and has been subsequently used in treating
non-critical variables such as in the random Potts model \cite{maria1}.
Concerning the Blume-Capel model defined in Eq. (\ref{eqn bcm}), 
this improved formalism provides now exact results for its
one-dimensional version and 
better values for the critical exponents and critical temperatures in
dimensions $d>1$ \cite{lara0} when compared to those from the previous 
usual application of the method \cite{osiel}. Besides, for the 
particular spin-1/2 case, the above formalism reproduces the early
results of the Ising model since $O_2=<S_i^2>=O_1=1/4$ is 
automatically satisfied.
 
The renormalization group Ansatz allows one to obtain, in addition,  an 
unambiguous estimate for
the non-critical thermodynamic function $O_2$ at criticality (note that the
mean field Ansatz gives {\it two} different estimates $O_2 = q$ and $O_2^\prime =
q^\prime$, respectively). However, it is not free from extracting results only
in plausible invariant subsets given by fixed ratios $r=K_2/K_1$. So,
complete analysis of the flux diagram is still not possible in such models.

\subsubsection{two uncoupled order-parameters}

Some systems present two different and independent order-parameters 
which can be used to go beyond the limitation (lacking of flux diagram) 
of the usual MFRG discussed so far.
For example, in the symmetric Ashkin-Teller model 
\cite{ashkinteller} where two different 
spin variables exist on each lattice site, it is possible to identify
two order-parameters in the problem. In a similar way, for the two-dimensional
Ising model with crossing bonds one can select appropriate pairs of 
order-parameters depending on the region of the parameter space. In such
models, for a finite block of $N$ spins, one has $O_1=O_1({\bf K},b_1)$ 
and $O_2=O_2({\bf K},b_2)$, where
$b_1$ and $b_2$ are the symmetry breaking fields related to the order-parameters
$O_1$ and $O_2$, respectively. Since for the uncoupled order-parameters
 $O_1 \rightarrow 0$ when 
$b_1\rightarrow 0$ (irrespective of the value of $b_2$) and
$O_2 \rightarrow 0$ when $b_2\rightarrow 0$ (irrespective of the value of $b_1$)
one has, to leading order in the $b's$,
\begin{equation}
 O_1=F_1({\bf K})b_1~,~~~~~~~~~
 O_2=F_2({\bf K})b_2~.
\label{eqn 26} 
\end{equation}
By doing the same calculations for a smaller block of $N^\prime$ sites and
assuming that both order-parameters are described by the same critical 
exponent (which is usually the case) one gets
\begin{equation}
F^\prime_1({\bf K^\prime})=F_1({\bf K})~, 
\label{eqn 27} 
\end{equation} 
\begin{equation}
F^\prime_2({\bf K^\prime})=F_2({\bf K})~.
\label{eqn 28} 
\end{equation}
So, contrary to the procedure prescribed in Eqs. (\ref{map1})-(\ref{map6}),
in this case
from equations (\ref{eqn 27}) and (\ref{eqn 28}) one can obtain renormalization
group flux diagrams of the usual type, furnishing isolated fixed points,
critical exponents, universality classes, etc. Quite good results 
(including some exact values
at some special points in the phase diagrams) have been obtained in the
Ashkin-Teller model \cite{oliveira2} and in the two-dimensional Ising
model with crossing bonds \cite{pla2} through the above scheme.

\subsubsection{two coupled order-parameters -- EMFRG}

The independence of the order-parameters, as stated in Eq. (\ref{eqn 26}),
is not commonly verified and there are some models where they do
not behave in such a simple form as discussed above. In the most general 
case one may have $O_1=O_1({\bf K},b_1,b_2)$ and $O_2=O_2({\bf K},b_1,b_2)$
so that, to leading order in $b_1$ and $b_2$, one gets
\begin{equation}
O_1=F_{11}({\bf K})b_1+F_{12}({\bf K})b_2~, 
\label{eqn 29} 
\end{equation}
\begin{equation}
O_2=F_{21}({\bf K})b_1+F_{22}({\bf K})b_2~,
\label{eqn 30} 
\end{equation}
where both order-parameters depend on $b_1$ and $b_2$. This general
problem can be tackled by defining the vectors ${\bf O}={O_1\choose O_2}$
and ${\bf b}={b_1\choose b_2}$
so that equations (\ref{eqn 29}) and (\ref{eqn 30}) can be written as
\begin{equation}
{\bf O}={\bf Fb}~,
\label{eqn 31} 
\end{equation}
where the matrix {\bf F} is given by
\begin{equation}
 {\bf F}=
\pmatrix{
F_{11}&F_{12}\cr
F_{21}&F_{22}\cr}~~.
\label{eqn 32} 
\end{equation}
Similar expressions are obtained by considering a finite system $N^\prime$
with less degrees of freedom. It is now easy to see that  when
the two order-parameters have the same critical exponents the corresponding
vectors ${\bf O}$ and ${\bf O^\prime}$ satisfy the scaling relation
\begin{equation}
{\bf O^\prime}=L^{d-y_H}{\bf O}~,
\label{eqn 32a}
\end{equation}
and the same for the symmetry breaking field vectors  
\begin{equation}
{\bf b}^\prime=L^{d-y_H}{\bf b}
\label{eqn 32b}
\end{equation}
if we take the usual MFRG. As a result, one ends up with the following matricial equation
\begin{equation}
({\bf F^\prime-F}){\bf b^\prime}=0~. 
\label{eqn 33} 
\end{equation}
The above matricial approach is  a natural extension to the general problem
of two-fold Hamiltonian parameters. This extended mean field renormalization
group (EMFRG) reproduces the previous cases as well as another
particular approach proposed more recently by Likos and Maritan 
\cite{likos}. Moreover, it can give the correct Bethe limit by using the
smallest clusters in treating the mixed spin Ising model (an early MFRG study
of the mixed spin-1/2--spin-1 \cite{verona} Ising model has failed to obtain the
corresponding Bethe results due to a different  mean field assumption in order to
decouple the order parameters). Some procedures in treating of the matricial Eq. 
(\ref{eqn 33})  are now in order:

(i) The simplest solution of Eq. (\ref{eqn 33}) is given by
\begin{equation}
det({\bf F^\prime-F})=0 ~.
\label{eqn 34} 
\end{equation}
From the above equation we do not have flow diagrams and the 
critical lines are obtained only for fixed values of $r=K_2/K_1$. 
This is a generalization of the corresponding one-order parameter 
where one has just
$F^\prime({\bf K^\prime})=F({\bf K})$. Now, by using the computed
expressions of reference \cite{verona} for the two order-parameters of the 
mixed spin-1/2--spin-1  Ising model one obtains the corresponding equation
(\ref{eqn 34}) which gives the same critical temperature from the Bethe
approximation by using the clusters depicted in Fig.1(a) and (b), as expected. 

(ii) If the two parameters are decoupled one has $F_{ij}=0, ~i\ne j$ for
any cluster and another solution for Eq. (\ref{eqn 33}) can be 
given by equating to zero all the matrix elements of ${\bf F^\prime-F}$, i.e.,
\begin{equation}
 F^\prime_{11}-F_{11}=0~{\rm and}~ 
F^\prime_{22}-F_{22}=0~,
\label{eqn 35} 
\end{equation}
which is the same result given by equations (\ref{eqn 27}) and (\ref{eqn 28}).

(iii) For the types of models considered by Likos and Maritan \cite{likos,likos2}
(in particular, models defined on a bipartite lattice with only two
different ground states, a ferromagnetic and an antiferromagnetic one, 
separated by a stable borderline)
one has $F_{11}=F_{22}$ and $F_{12}=F_{21}$ for any finite system.
The last two relations  reflect
the additional symmetry $O_1({\bf K},b_1,b_2)=O_2({\bf K},b_2,b_1)$ exhibited
by these order parameters. So, by 
equalling again to zero all the matrix elements of ${\bf F^\prime-F}$ one gets
\begin{equation}
F^\prime _{11}-F_{11}=0~{\rm and} ~ F^\prime _{12}-F_{12}=0~,
\label{eqn 36} 
\end{equation}
which are the same expressions as those proposed in reference \cite{likos}
in a different context.

One can thus see that the present matricial EMFRG  is capable to 
generalize all the previous isolated treatments done on  two-dimensional
parameter Hamiltonians. One can also notice that: 

({\it a}) in
the general case where all the matrix elements of ${\bf F}$ are different
from each other, Eq. (\ref{eqn 33}) can furnish four distinct recursion
relations, i.e., at first sight more equations than parameters. However, it should
be possible, in these cases, to augment the original Hamiltonian by a
suitably chosen number of interactions until the number of couplings
matches the number of recursion relations. From the flow diagram
so obtained in this enlarged space one could get the corresponding flows
in the original restricted domain by taking the appropriate subspace in
the complete Hamiltonian space. 

({\it b}) in addition, the present formalism
can also be straightforwardly applied to models having more than
two order-parameters simply by computing the dimensional enhanced vectors ${\bf O}$
and ${\bf b}$ and the corresponding matrix ${\bf F}$ for any finite cluster.

Unfortunately, the extensions ({\it a}) and ({\it b}) above
have not yet been tested in studying statistical mechanics systems from MFRG
and one can not say, a priori, that this approach will succeed in giving
the correct critical behavior in these cases, as is usual in ordinary real space 
renormalization group procedures.

\subsection{Dynamic MFRG (DMFRG)}

So far, just equilibrium static properties have been treated through MFRG.
In this subsection we discuss the extension of the mean field renormalization
group ideas to non-equilibrium phenomena. Despite the simplicity
of the formalism, the MFRG has been applied to a very few number of systems 
out of the thermodynamic equilibrium. It was employed for the first time
by Indekeu, Stella and Zhang \cite{zhang} to study the dynamics of the 
kinetic Ising model with single spin-flip Glauber transitions \cite{Glauber} 
near the equilibrium states. It has also been used to treat the 
dynamic critical properties of quantum spin systems \cite{pla3}.
As the extensions to treat the dynamics of quantum models is (at least
in principle) straightforward to be done,
we present below just the formalism for treating kinetic classical models with
special emphasis to the Ising model by considering larger clusters
than in the original work of Indekeu et al \cite{zhang}. Extensions to other
problems are rather easy to be performed and will be summarised in section V.

In the dynamic approach of the MFRG one starts from the scaling relation of
the magnetization, close to equilibrium and criticality. For sufficiently
long times $t$ we expect the magnetization of the infinite system to scale as
\cite{ma}
\begin{equation}
m(\epsilon ,H,t) = \ell^{-d+y_{H}}m(\ell^{1/\nu}\epsilon ,
           \ell^{y_{H}}H,\ell^{-z}t),
\label{eqn 37}
\end{equation}
which is a generalization of Eq. (\ref{eqn 0bb}) where $z$ 
is the corresponding dynamic critical exponent.

Let
${\cal S}=(S_{1},S_{2},\ldots,S_i,\ldots,S_{N})$, with $S_{j}=\pm 1$, 
represent a state of the finite system with $N$ spins, and $P({\cal S},t)$ the 
probability of finding the system in the state  ${\cal S}$ at instant $t$. 
The time evolution of  $P({\cal S} ,t)$ is given by the master equation
\begin{equation}
{dP({\cal S},t)\over dt} = \sum_{i=1}^{N} [P({\cal S}^{i},t)W_{i}({\cal S}^{i}) - 
P({\cal S},t)W_{i}({\cal S})]~,
\label{eqn 38}
\end{equation}
where $W_{i}({\cal S})$ is the transition rate, per unit time, to flip the 
spin $S_{i}$, and
${\cal S}^{i}=(S_{1},S_{2},\ldots,-S_{i},\ldots,S_{N})$.
The first sum in the above equation takes into account all transitions to the
state ${\cal S}$ and the second sum all the transitions out from
the state ${\cal S}$.
If $f({\cal S})$ is a given function of state ${\cal S}$, we can evaluate its 
average value by 
\begin{equation}
<f({\cal S})> = \sum_{{\cal S}^\prime}f({\cal S}^\prime)P({\cal S}^\prime,t)~,
\label{eqn 39}
\end{equation}
and its time evolution through the expression 
\begin{equation}
{d<f({\cal S})>\over dt} = \sum_{i=1}^{N}<[f({\cal S}^{i})-f({\cal S})]W_{i}
({\cal S})>~.
\label{eqn 40}
\end{equation}
By taking the function of state  as being the local magnetization
of the spin $i$, we can easily write
\begin{equation}
{d<S_{i}>\over dt} = -2<S_{i}W_{i}({\cal S})>~.
\label{eqn 41}
\end{equation}

Let $m_{N}(t)$ and $m_{N^\prime}(t^\prime)$ be the time dependent
magnetizations of two finite
clusters with N and $N^\prime$ spins, respectively, i.e.,
\begin{eqnarray}
	m_{N}(t)        & = & {1\over {N}} \sum_{i=1}^{N}<S_{i}> , 
	\label{eqn 42}\\
	m_{N^\prime}(t) & = & {1\over {N^\prime}} \sum_{i=1}^{N^\prime}
<S_{i}^\prime>~,
        \label{eqn 43}
\end{eqnarray}
where the time dependence comes from $<S_i>$ through Eq. (\ref{eqn 41}).
Following the original idea of the MFRG , each spin $S_{i}$ of the 
border of the cluster couples with an infinitesimal symmetry breaking field
$b(t)$. The equations of motion for $m_{N}(t)$ and $m_{N^\prime}(t^\prime)$
are computed from Eqs. (\ref{eqn 39})-(\ref{eqn 41}) by using
\cite{Glauber}
\begin{equation}
W_{i}({\cal S}) = {1\over 2} [1 - S_{i} \tanh(K\sum_{j\not=i}
S_{j} + c_i K b_{N})]~,
\label{eqn 46}
\end{equation}
where $c_i$ is the number of first neighbors of  $S_{i}$, which are 
external to the cluster, and 
\begin{eqnarray}
P_{N}({\cal S},t) & = & {1\over 2^{N}}(1 + \sum_{i}m_{i}(t)S_{i} + \sum
_{{i,j}
\atop{ \mbox{$i\neq j$}}} p_{ij}(t)S_{i}S_{j} + \nonumber \\
                &   & \sum_{{i,j,k}\atop{ \mbox{$i\neq j\neq k$}}}q_{ijk}(t)
S_{i}S_{j}S_{k} + \ldots)~,
\label{eqn 47}
\end{eqnarray}
where $m_{i}(t)=<S_{i}>$, $p_{ij}(t)=<S_{i}S_{j}>$, 
$q_{ijk}(t)=<S_{i}S_{j}S_k>$, etc.
The transition rate $W_{i}({\cal S})$ is chosen to satisfy the detailed balance
condition $P({\cal S}^{i},t)W_{i}({\cal S}^{i}) = P({\cal S},t)W_{i}({\cal S})$
in the limit of $t \rightarrow \infty $. This is a sufficient condition
to the equilibrium be attained, i.e., $dP({\cal S},t)/dt=0$ as $t \rightarrow 
\infty $ with the expected probability distribution
$P_{N}({\cal S},t\rightarrow \infty) \propto e^{-\beta E({\cal S})}$, where
$E({\cal S})$ is the energy of the state ${\cal S}$. The expression for the
probability $P_{N}({\cal S},t)$, Eq. (\ref{eqn 47}), is an exact one and can
be worked out just for rather small systems.

In general, for small clusters at $H=0$ and linearized in  
$b_{N}(t)$ and $b_{N^\prime}(t^\prime)$ the time derivatives of the
order parameters can be written as
\begin{eqnarray}
dm_{N}(t)\over dt        & = & -A_{N}(K)m_{N}(t) + B_{N}(K) b_{N}(t)~,
\label{eqn 44} \\
dm_{N^\prime}(t^\prime)\over dt^\prime & = & -A_{N^\prime}(K^\prime)m_{N^
\prime}(t^\prime) + B_{N^\prime}(K^\prime) b_{N^\prime}(t^\prime)~,
\label{eqn 45}
\end{eqnarray}
where the expressions for  $A_{N}(K)$, $A_{N^\prime}(K^\prime)$, $B_{N}
(K)$ and $B_{N^\prime}(K^\prime)$ are computed from Eqs. 
(\ref{eqn 39})-(\ref{eqn 47}). 

According to the strategy of MFRG, we impose now the following scaling 
relations for  the magnetizations and symmetry breaking fields:
\begin{equation}
m_{N^\prime}(K^\prime,0,\ell^{-z}t;
b_{N^\prime}(\ell^{-z}t)) = \ell^{d-y_{H}}m_{N}(K,0,t;b_{N}(t))~,
\label{eqn 48}
\end{equation}
and
\begin{equation}
b_{N^\prime}(\ell^{-z}t) = \ell^{d-y_{H}}b_{N}(t)~,
\label{eqn 49}
\end{equation}
where $\ell=({N\over{N^{\prime}}})^{{1\over {d}}}$
and $t^\prime=\ell^{-z}t$ with $N^{\prime} < {N}$.

Taking the derivative of Eq. (\ref{eqn 48}) with respect to $t$, and using Eqs.
(\ref{eqn 44}), (\ref{eqn 45}) and (\ref{eqn 49}), we arrive at the following 
recursion relations  for the $A's$ and $B's$ coefficients:
\begin{eqnarray}
	A_{N^\prime}(K^\prime) & = & \ell^{z}A_{N}(K)~, 
	\label{eqn 50} \\
	B_{N^\prime}(K^\prime) & = & \ell^{z}B_{N}(K)~.
	\label{eqn 51}
\end{eqnarray}
The solution of the above system of equations provides the non-trivial 
fixed point 
$(K=K^\prime=K_{c})$ and furnishes also the value of the dynamic critical 
exponent $z$. 

As a simple example, let us take the one and two spin clusters
depicted in Fig. 1(a) an (b) for this dynamic treatment in the $d$-dimensional 
Ising model. We have then
\begin{equation}
{dm_{1}(t)\over dt}=-2\sum_{\{S_1\}}S_1{1\over 2}[1-S_{1}\tanh(cK^\prime b_{1})]
{1\over 2}[1+m_1S_1]~,
\label{eqn 51a}
\end{equation}
\begin{eqnarray}
{dm_{2}(t)\over dt }= -2 \sum_{\{S_1,S_2\}}S_1{1\over 2}[1 - S_{1} 
\tanh(KS_2+(c-1) K b_{2})]\times \nonumber \\
\times {1\over 4}[1+m_2(t)S_1+m_2(t)S_2+p_{12}(t)S_1S_2]~,
\label{eqn 51b}
\end{eqnarray}
where in the latter relation we have used the fact that both spins in the
two site cluster of Fig. 1(b) have the same time dependence. Performing the sum
over the corresponding states in the two relations above one gets
\begin{equation}
{dm_{1}(t)\over dt} = -m_1(t) + cK^\prime b_1(t)~,
\label{eqn 51c}
\end{equation}
\begin{eqnarray}
{dm_{2}(t)\over dt }= -(1-\tanh K)m_2(t) + (c-1)K(1-\tanh ^2 K)b_2(t)~,
\label{eqn 51d}
\end{eqnarray}
from which one has
\begin{eqnarray}
	1 & = & \ell^{z}(1-\tanh K)~, 
	\label{eqn 51e} \\
	cK^\prime & = & \ell^{z}(c-1)K(1-\tanh ^2 K)~.
	\label{eqn 51f}
\end{eqnarray}
By taking the ratio of Eq. (\ref{eqn 51f}) to Eq. (\ref{eqn 51e}) it is
easy to see that 
the non-trivial fixed point and the related thermal critical exponent
are the same as those obtained in the static procedure of section IIA
(as well as the magnetic critical exponent $y_H$ if one includes the external
field $H$), as expected. Estimates of the corresponding dynamic 
critical exponent $z$ are also readily achieved from above equations
(see Table \ref{tab 4}).

In   Table \ref{tab 4}  we show the results obtained in the study
of the Ising model for clusters of
up to  $9$ spins in two dimensions (a cluster with non-equivalent
boundary sites which has not been previously considered  in the
dynamic treatment of the model) and with up to $8$ spins in three 
dimensions. One can see from this Table that the results are
consistent with those from other approaches. However, some difficulties
arise in considering bigger clusters in the present case:

(a) when we consider a cluster with four  spins, by using Eq. (\ref{eqn 47})
it appears correlations between
pairs of spins which are located at the diagonals of the  cluster;
 
(b) we must also be careful in the calculation of the mean values $<S_i>$,
because for clusters with more than four spins they are not all equivalent. 
It should be stressed that this non-equivalence among spins 
inside the finite cluster for
$N>4$ in two-dimensions (and $N>8$ in three-dimensions) implies
in the necessity to diagonalize a system of equations in order to 
obtain an expression like that of Eq.(\ref{eqn 44}), since in this case $m_N(t)$
is no longer an eigenmode of the dynamics but is coupled
to higher spin correlations; 

(c) besides, another great 
difficulty arises when we try to increase  the size of the  clusters in this 
non-equilibrium approach of the MFRG. Although the method does not produce 
couplings of longer range in space, it is necessary to take account of the 
higher than two spin correlations, which naturally appear for bigger clusters. 
Then, we need to use methods of equilibrium statistical mechanics to 
evaluate these higher order correlations, although we are interested  only in
the determination of the equation of motion for the mean magnetization of 
the cluster.
 
As a result, the  dynamic MFRG must be applied  only to
small clusters.

\section{Exact results from MFRG}

Besides the great qualitatively success achieved by applying the MFRG to a great variety
of statistical systems (even by taking the simplest choice for the clusters)
some expected rigorous quantities are obtained in some particular cases, as we 
have already seen in the previous section. That
is what happens, for instance, when treating the one-dimensional spin-1/2
Ising model \cite{indekeu}, the two-dimensional spin-1/2 isotropic Heisenberg
model \cite{pla4} and the spin-S one-dimensional Blume-Capel model 
\cite{lara0}. While the exact critical temperature ($T_c=0$) is 
reproduced in all models above for clusters of any size, in some of them the
exact critical exponent is only obtained in the limit of infinite cluster
(except for the magnetic exponent  $y_H$ in the Ising model and the
thermal exponent $\nu$ in the $d=2$ Heisenberg model, where they
are exact even for the smallest clusters).

One also obtains expected quantitative values at some special points in the
global phase diagram of the Ashkin-Teller model \cite{pla1,oliveira2} and
the Ising model with crossing bonds \cite{pla2,likos}. 
Exact results are
also achieved by applying the method in the study of the geometrical problem
of directed self-avoiding random walk in two dimensions and the Ising model on
the Bethe lattice \cite{neves}. These rigorous results reproduced by the
MFRG in so many different systems such as classical, quantum and geometrical 
statistical problems assign indeed to the method  a high reliability 
when studying more complex models.

\section{ Static Problems Applications}

The previous two sections give us an idea of the performance 
and potentiality of the phenomenological 
mean field renormalization group when used to treat some particular models. 
In this and in the next  section we will present, in a summarized way, the applications 
of the method (in its various forms) to several  different statistical mechanical systems.
It is not the purpose here to discuss the results obtained for every studied system
(which amounts to over 120 papers).
Rather, just an update of the treated models by employing the method 
will be presented. Moreover, this section is intended to be self-contained in the
sense that all the references already commented in the text is referenced here
again when listing the corresponding model where it has been used.

The Ising model has been extensively studied and has also been a 
test system for all  proposed procedures involving the MFRG approach
\cite{indekeu,croes,indekeu2,slotte,indekeu3,pla5,shi,peli}, 
including a combination with the decimation procedure \cite{osiel2,tang},
the effects of reaction \cite{droz3} and symmetry dependent fields \cite{kamien},
and  treatments of some extended models such as:
antiferromagnetic model on triangular  \cite{slotte2}
and square lattices with external field \cite{slotte3};
anisotropic Ising model on a triangular lattice\cite{likos2};
surface effects in pure and random semi-infinite models 
\cite{marques0a,santos1,jiang1,jiang2,xu,moreira,ilkovic1,li,gang1,gang2,arruda0}; 
the Ising system on a compressible lattice (Domb model) \cite{pla6}; 
and the model with crossing bonds in two-dimensions \cite{pla2,likos,pla7}.
Additional diluted versions of the Ising model have been treated by considering the
effects of different kinds of random dilution including spin-glass
\cite{droz,droz3,evan1,evan2,yang3,benayad,rosales,pla8,douglas,mina,ricar,bohor,zamora,ricar2,zamora2,douglas2},
dilution in the antiferromagnetic model in an external field \cite{reiner} 
and random field systems \cite{droz,arruda}. 

Geometrical critical phenomena and percolation have been studied by
De'Bell \cite{bell1,bell2} and das Neves and Kamphorst Leal da Silva
\cite{neves2}, and some
exact results have been achieved \cite{neves}. The method
has also been used in the study of directed percolation hypothesis for 
stochastic cellular automata \cite{odor}, lattice gas model \cite{surda}, and
nematic ordered states at low temperatures \cite{leitao,leitao2}.

Other related systems include: Ashkin-Teller model \cite{pla1,oliveira2,pawli};
coupled spin-1/2 Ising models \cite{cesare}; ANNNI model \cite{valadares}; 
mixed spin-1/2 spin-1 model \cite{verona,samaj}; different versions of
pure and random Potts models on infinite and  semi-infinite lattices
\cite{maria1,marques00b,santos2,marques0b,marques0c,ye,marques0d,marques0e}; 
the $Z_4$ spin model \cite{riera}; Hamiltonian version of the
two-dimensional $Z(q)$ symmetric spin models \cite{niebur,niebur2};
quenched and annealed random-bond D-vector models \cite{lyra}; 
the planar random anisotropic model \cite{mucio}; $q$-state clock spin-glass
models \cite{chen}; classical
\cite{osiel,lara0,ilkovic2,wagner,lara1} and quantum random versions
of the Blume-Capel model \cite{ssantos}; the Blume-Emmery-Griffiths model
\cite {osiel3}; the finite temperature
SU(2) lattice gauge theory at the strong coupling limit \cite{gunduc};
and classical XY and Heisenberg models \cite{ricar2b}.

A great variety of quantum spin systems have already been studied from the 
MFRG approach. Among them we have: anisotropic Heisenberg model 
\cite{pla4,bukman,ricar3,ricar3a,ricar3b}; anisotropic Heisenberg model 
in a transverse field  \cite{cao}; pure and random 
spin-1/2  \cite{pla0,valadares2,yang} and spin$>$1/2 
\cite{song1,song2,song3,song4}  transverse Ising models;
quantum Ising model with annealed antiferromagnetic bond randomness
\cite{shi2}; Ising model in random transverse field 
\cite{song4,cassol,song5,song6,serkov1,song7};
random mixture of Ising and Heisenberg models \cite{droz2,martins};
spin-1 anisotropic Heisenberg chain \cite{solyom} 
and spin-1 Ising model in a transverse field \cite{yang2};
quantum models applied to granular superconductors \cite{granato}; and
anisotropic Heisenberg model with Dzyaloshinskii-Moryia  interaction
\cite{serkov2}.

More recent extensions of the method include the study of the MFRG approach
with the Tsallis Statistics \cite{ricar4} and non ambiguous location of 
multicritical points \cite{lara1}.

\section{Dynamic Problems Applications}

Although fewer in number, the DMFRG has been applied in the study of
dynamics and non-equilibrium properties of some interesting systems. 
For instance, the classical approach of Indekeu et al \cite{zhang}
has been extended in the treatment of the intrinsic dynamic critical properties
of quantum spin models \cite{pla3}. In this case, good results for the
critical exponent $z$ have been obtained when compared to others 
from more laborious approaches. 

It was also applied to non-equilibrium problems arising
from the competition between  microscopic mechanisms. 
This is the case, for instance, of the Ising model 
subject to two locally competing temperatures \cite{Marques1,Tania}. In this
model, for each temperature, the corresponding transition rate satisfies
the principle of detailed balance. When the two processes are considered
at the same time, a continuous phase transition is observed between steady
states. By using clusters of one and two spins Marques \cite{Marques1} was able to
determine the phase diagram of the model in the temperature
versus gradient of temperature plane. She also obtained values for the critical
exponent $\nu$ of this non-equilibrium model in two and three dimensions,
and they  compare very well with the values  of the corresponding equilibrium
Ising model. On the other hand, if one of the temperatures becomes
negative, the ferromagnetic system can display an antiferromagnetic order
\cite{Tania}. The heat bath at negative temperature simulates a flux of
energy into the system. The MFRG was also applied to other non-equilibrium
models which exhibit steady states: non-equilibrium Ising model with 
competing dynamics \cite{marques1a},
critical behavior of non-equilibrium 3-state systems
\cite{Ceu}, stochastic cellular automata in one and two dimensions \cite{odor}, 
competition between diffusion and
anihilation of many particles \cite{Marques2}, dilution in the contact
process \cite{Marques3}, critical surface behavior of models with one 
absorbing state \cite{Marques4} and one-dimensional models with  multiple 
absorbing configurations \cite{Mendes}.

This methodology has also been used to study the dynamic critical behavior of
a semi-infinite system of spins in a simple  cubic lattice 
with nearest-neighbor ferromagnetic interactions. As this approach 
has not been published in the literature and presents some similarities
with the one discussed in section II.C.3 we outline  below
the general procedure. It is assumed that in the surface plane of the 
semi-infinite system the exchange coupling is given by  $K_{S}$, while all 
the other couplings are given by $K$. For $\rho =K_S/K$ greater than a value
$\rho _c$ the surface orders before the bulk  while for  $\rho < \rho _c$
the surface orders when the bulk does. As an example, in the one-site cluster 
of the MFRG, we take one spin in the $i-th$ plane, and it interacts with the
symmetry breaking fields: $b^{i}_1(t)$ of 
the $i-th$ plane, and $b^{i-1}_i(t)$ and $b^{i+1}_1(t)$ of the adjacent planes.
Here, and from now on, the superscript designates the particular plane 
away from the $i=1$ surface.
As is usual, these fields are assumed to be very small, that is, we are 
considering the problem in the neighborhood of the surface phase transition.
The equation of motion for the magnetization taking the one-site cluster, in each  
plane, is then given by 
\begin{equation}
{dm^1_1(t)\over dt } =  -m^1_1(t) + 4K_{S}^\prime b^1_{1}(t)+K^\prime b^{2}_1(t)~,
\label{eqn 52}
\end{equation}
for the magnetization on the surface $i=1$ and, for $i\geq 2$,
\begin{equation}
{dm^i_{1}(t)\over dt}  =  -m^i_{1}(t) +
K^\prime [b^{i-1}_1(t)+4b^{i}_1(t)+b^{i+1}_1(t)]~.
\label{eqn 53}
\end{equation}
Eq. (\ref{eqn 53}) is a generalization of Eq. (\ref{eqn 51c}) for the bulk
magnetization with different symmetry breaking fields in different planes
and Eq. (\ref{eqn 52}) is a particular case of relation (\ref{eqn 53}) for
the surface magnetization with intra-plane interactions $K_S$.
Similarly, the generalized expressions for the two spin cluster contained
in each plane read
\begin{equation}
{dm^1_2(t)\over dt } =  -(1-\tanh K_S)m^1_2(t) + 
(1-\tanh ^2K_S) [3K_{S}b^1_{2}(t)+K b^{2}_2(t)]~,
\label{eqn 52a}
\end{equation}
\begin{equation}
{dm^i_2(t)\over dt } =  -(1-\tanh K)m^i_2(t) + 
(1-\tanh ^2 K) [Kb^{i-1}_2+3Kb^i_{2}(t)+K b^{i+1}_2(t)]~.
\label{eqn 53a}
\end{equation}
In general, the magnetization for the planes $i$ taking a small block of
$N$ spins inside each layer can be written as
\begin{equation}
{dm^1_{N}(t)\over dt} = -A^1_{N}(K_S)m^1_{N}(t) + 
\sum_{j=1}^{2}B^j_{N}(K,K_S)b^j_{N}(t)~,
\label{eqn 54}
\end{equation}
\begin{equation}
{dm^i_{N}(t)\over dt} = -A^i_{N}(K)m^i_{N}(t) + 
\sum_{j=i-1}^{i+1}B^j_{N}(K)b^j_{N}(t)~,~~i\ge 2~,
\label{eqn 54i}
\end{equation}
where the explicit dependence of the coefficients $A^i_{N}$
and $B^i_{N}$ on $K$ and $K_S$ has been given. 

By considering now two clusters of $N$ and $N^\prime$ spins and the scaling
relations given by equations (\ref{eqn 48}) and (\ref{eqn 49}) for
the magnetizations on each plane (and introducing adequately the
exponents $y_{Hs}$ and $z_S$ for the surface magnetization and $y_H$
and $z$ for the bulk quantities) one finds, after taking the
derivative with respect to the time $t$, a set of linear equations relating
the magnetizations and the corresponding symmetry 
breaking fields of different layers.
The set of linear equations for the magnetizations can be written as
\begin{equation}
{\bf \cal A}_N {\bf \cal M}_N = {\bf \cal A}_{N^\prime} {\bf \cal M}_{N^\prime} ~,
\label{eqn 54a}
\end{equation}
where ${\bf \cal  M}_N$ is a column vector composed by the layer
magnetizations $m^1_N,~m^2_N,~m^3_N,~...,$ and ${\bf \cal A}_N$ is a
diagonal matrix whose elements are straight related to $A^i_{N}$. The
same holds for the smaller system. A solution of (\ref{eqn 54a}) is
obtained by equating all the terms of the diagonal of the matrices and,
as all terms for $i\ge 2$ are the same, results basically in
\begin{equation}
A^1_N(K_S)=\ell ^{-z_S}A^1_{N^\prime}(K^\prime _S)~,~~~
A^i_N(K)=\ell ^{-z}A^i_{N^\prime}(K^\prime)~,
\label{eqn 54b}
\end{equation}
which gives a relation among $K_S$, $K$, $z$ and $z_S$ at criticality. 
The set of linear equations for the remaining symmetry breaking
fields can be put in the following form
\begin{equation}
{\bf {\cal B}  b}=0~,
\label{eqn 54c}
\end{equation}
where ${\bf  b}$ is the column vector composed by the symmetry breaking
fields $b^1_N,~b^2_N,~b^3_N,~...,$ and ${\bf \cal B}$ is a three-diagonal
matrix whose elements, using relations (\ref{eqn 54b}), are just functions of
$K_S$ and $K$ at criticality. It is easy to see that the critical coupling to 
give an ordered surface  over a paramagnetic bulk phase is found when the 
determinant of matrix  ${\bf \cal B}$ vanishes. One then
obtains an equation for the surface critical temperature as a function
of $\rho =K_S/K$, as is usual in the MFRG approach of two parameter systems.
The previous bulk dynamic properties are reproduced, as expected, for $i\gg 1$.

For the simplest finite systems of Fig. 1 it is not difficult to show
from Eqs. (\ref{eqn 52})-(\ref{eqn 53a}) and (\ref{eqn 54a})-(\ref{eqn 54c}) that
\begin{equation}
{\tanh K_{Sc} \over {1 - 3 tanh K_{Sc}}}=\rho~,
\label{eqn 54d}
\end{equation}
for the surface critical temperature $K_{Sc}$ as a function of the ratio $\rho$ and 
\begin{equation}
1 - tanh Ks = l^{-zs}~,  ~~~~~1 - tanh K = l^{-z}~,
\label{eqn 54e}
\end{equation}
from which the dynamic critical exponents are obtained. The 
numerical results from above equations are shown
in Table \ref{tab 5} together with those obtained by taking the 
plaquette of four spins. We can note that the DMFRG is easily extended
to study the critical dynamic phenomena on surfaces and, in particular,
furnishes a value for $\rho _c =1.48$ comparable to that from
Monte Carlo simulations $\rho _c =1.52$ \cite{Landau}, while the mean field result
is $\rho_c=1.25$ \cite{binder,hohenberg}. It should also be noticed that in the present approach with
rather small clusters the surface magnetic exponent $y_{HS}$ is 
indeterminate since it cancels out in the renormalization group equations.

\section{Related Phenomenological RG}

Different phenomenological renormalization group procedures based on exact 
calculations in finite systems have also been proposed in the literature. 
In this section some of them will be discussed, namely, the effective field 
renormalization and finite size scaling renormalization groups. 
Few other methods, which are less used  or applied in approaches other 
than the renormalization group
scheme, will be referenced in the next section. Here, as in the MFRG case, 
we are mainly concerned in describing the methods themselves and their 
interrelations and similarities, and just a list of their 
applications in different statistical models will be presented.

\subsection{Effective Field Renormalization Group (EFRG)}

The MFRG in its several formulations computes exactly the magnetization 
(order parameter) for each finite cluster of spins according to the
canonical distribution (\ref{eqn 5a}). An alternative way of obtaining  
this order parameter for a Hamiltonian model system ${\cal H}$
has been proposed by some authors \cite{li2,benayad2,fitti}
by employing an effective field theory based on the exact generalized 
Callen-Suzuki identity \cite{callen,suzuki}
\begin{equation} 
\left< O_n \right> = \left< { {Tr_nO_n\exp {-\beta\cal H}_n}\over
                     {Tr_n\exp {-\beta \cal H}_n}    }\right>~,
\label{eqn 55} 
\end{equation}
where the partial trace $Tr$ is taken over the set of $n$ spins variables
specified by a finite system Hamiltonian ${\cal H}_n$, $O_n$ is the 
corresponding order parameter (or, in general, any other function of all the $n$ 
spins of the cluster) and $<\cdot \cdot \cdot>$ indicates the usual canonical 
thermal average taken over the ensemble defined by the complete 
Hamiltonian ${\cal H}$. The idea is simply to replace  ${\cal H}$ by
a cluster of $n=N$ spins surrounded by fixed magnetizations at values
$b$. In this way the order parameter $m=O$ in Eq. (\ref{eqn 55})
can be computed by employing the 
exponential operator technique \cite{homura} resulting, for $b<<1$, 
in an equation of a similar form as before for $H=0$, namely,
\begin{equation}
m_N=f_N({\bf K})b_N~. 
\label{eqn 56} 
\end{equation}
Doing the same for a smaller system ($N^\prime <N$) and using the
mean field renormalization group assumptions given by
equations (\ref{eqn 6}) and (\ref{eqn 10}) one obtains the EFRG recursion
relation
\begin{equation}
f_{N^\prime} ({\bf K^\prime})=f_N({\bf K})~, 
\label{eqn 57} 
\end{equation}
from which phase diagrams and estimates of critical exponents are computed.

In order to illustrate the method let us consider the spin-1/2 Ising model
and $n=1$ in Eq. (\ref{eqn 55}). The Callen-Suzuki identity 
(\ref{eqn 55}) then reads
\begin{equation}
m_1=<S_1>=\left<\tanh (K^\prime\sum_{j=1}^cS_j)\right>~,
\label{eqn 57a}
\end{equation}
where the sum is over all nearest neighbors spins of $S_1$. By using the 
exponential operator technique \cite{homura}
\begin{equation}
e^{\delta D_x}f(x)=f(x+\delta ),~~~~~~~~~D_x={\partial \over {\partial x}}~,
\label{eqn 57b}
\end{equation}
the Eq. (\ref{eqn 57a}) can be written as
\begin{equation}
m_1=\left< e^{\delta D_x} \tanh (x) \mid _{x=0}\right>~,~~~~~~~~~
\delta =K^\prime\sum_{j=1}^cS_j~,
\label{eqn 57c}
\end{equation}
and, as the hyperbolic tangent does not depend on any spin configuration one has
\begin{equation}
m_1=\left< \prod _{j=1}^ce^{K^\prime S_jD_x}\right>\tanh (x) \mid _{x=0}=
\left< \prod _{j=1}^c[\cosh ({K^\prime D_x})+S_j\sinh ({K^\prime D_x})]\right>\tanh x \mid _{x=0}~,
\label{eqn 57d}
\end{equation}
where in the last expression we have used the van der Waerden identity for 
the two state system 
\begin{equation}
e^{aS_i}=\cosh (a) + S_i\sinh (a),~~S_i=\pm 1~.
\label{eqn 57dd}
\end{equation}
The above equations are still exact relations and computation of $m_1$
will commonly require some approximations such as decoupling the mean values of
product of spins in product of spin mean values \cite{homura}.
We can also see from Eq. (\ref{eqn 57d}) that the mean value $m_1$ is strongly
dependent on the number of first neighbors $c$. By taking the one spin
cluster $N^\prime =1$  of Fig. 1(a) we see that $S_j=b_1$ for all $j$.
The order-parameter for the one-dimensional lattice
Eq. (\ref{eqn 57d}) assumes then the form
\begin{equation}
m_1=\left<[\cosh (K^\prime D_x) + b_1 \sinh (K^\prime D_x)]^2\right>
\tanh (x) \mid _{x=0}~.
\label{eqn 57e}
\end{equation}
Since $\sinh (K^\prime D_x) \cosh (K^\prime D_x) \tanh (x) | _{x=0} = 
{1\over 2} \tanh (2K^\prime)$,
$\sinh ^2 (K^\prime D_x) 
\tanh (x) | _{x=0} =0$ \hfill \break and $\cosh ^2 (K^\prime D_x) \tanh (x) 
| _{x=0} = 0$ 
one has for a homogeneous system
\begin{equation}
m_1=\tanh (2K^\prime )~ b_1~.
\label{eqn 57f}
\end{equation}
It is interesting to note that the usual mean field approach $m_1=b_1$ gives
the exact result $T_c=0$ even for the one spin cluster. In fact, this
approach reproduces all the expected results for the thermodynamic
properties, such as magnetic susceptibility and
specific heat, of the linear chain in zero external field
\cite{cesar} and reflects the fact that, besides appearing at most just
pair correlation functions in the thermodynamic functions, the 
auto-correlation $<S_i^2>=1$ has been taken into account exactly 
in the van der Waerden  relation (\ref{eqn 57dd}). Analogously, 
for the two-dimensional square lattice one gets \cite{homura}
\begin{equation}
m_1={1\over 2}[\tanh(4K) + 2\tanh (2K)]b_1~,
\label{eqn 57g}
\end{equation}
which furnishes the mean field critical temperature $K_c=0.324$, and for
the three-dimensional cubic lattice 
\begin{equation}
m_1={3\over {2^4}}[\tanh (6K^\prime)+4\tanh (4K^\prime)+5\tanh (2K^\prime)]~b_1,
\label{eqn 57h}
\end{equation}
giving $K_c= 0.197$ \cite{homura}. As usual, from the above simple approach 
one always obtains classical critical exponents (as can be seen by expanding
further up to the order $b_1^3$).

For a two spin cluster $n=2$  Eq. (\ref{eqn 55}) reduces to \cite{bobak}
\begin{equation}
m_2={1\over 2}\left<S_1 + S_2\right>=
     \left<\prod _{j=1}^{c-1}e^{KS_jD_x}
          \prod _{j^\prime =1}^{c-1}e^{KS_{j^\prime}D_y}\right>
          {\sinh (x+y) \over {\cosh (x+y) +e^{-2K}\cosh x-y)}}~,
\label{eqn 57i}
\end{equation}
where $D_y = {\partial \over {\partial y}}$ and the sum in $j$ is over
the first neighbors of the spin $1$, and the sum in $j^\prime $ is over
the first neighbors of the spin $2$. It is now easy, though rather lengthy,
to compute the expressions for $m_2$ taking the cluster
of Fig. 1(b) on different lattices. We quote below 
just the results for the one- and two(square)-dimensional lattices
which are, respectively,
\begin{equation}
m_2={f(2)\over {g(2) + e^{-2K}}}~b_2~,
\label{eqn 57j}
\end{equation}
\begin{equation}
m_2={3\over {2^4}}\left({{f(6) \over {g(6)+e^{-2K}}} +{4f(4) \over {g(4)+e^{-2K}g(2)}}+
{2f(2) \over {g(2)+e^{-2K}g(4)}} + {3f(2) \over {g(2)+e^{-2K}}}}\right)b_2~,
\label{eqn 57hh}
\end{equation}
where $f(u)=\sinh uK$ and $g(u)=\cosh uK$. A complete set of 
coefficients for other types of lattices taking the smallest clusters
can be found in reference \cite{fitti}. Eq. (\ref{eqn 57j}) gives also
$T_c=0$ for the one-dimensional case and one gets $K_c=0.331$ and
$K_c=0.198$ for $d=2$ and $d=3$, respectively \cite{bobak}.

The EFRG obtained from Eqs. (\ref{eqn 56}),
(\ref{eqn 57}), (\ref{eqn 57f}) and (\ref{eqn 57j}) 
reproduce the exact one-dimensional
results, including the expected thermal critical exponent $\nu$ even for
the smallest systems.
In general, the corresponding renormalization recursion relations 
for dimensions $d>1$ give better values than the MFRG, as can be seen in
Table \ref{tab 6}, where a comparison is made on square and simple cubic Ising models.
In addition, the two spin block is, within this formalism, able to distinguish
the two dimensional triangular lattice from the simple cubic one
(both with the same coordination number $c=6$) regarding the critical
temperature ($K_c=0.244$ for the triangular lattice and the exact value
is $K_c=0.275$ \cite{baxter}). This distinction comes
from the fact that on a triangular lattice the total number of surrounding
mean field sites for $N=2$ cluster is $8$ and in the simple 
cubic one is $10$. While 
Eq. (\ref{eqn 57i}) depends on this value, in the MFRG approach 
expressed by Eq. (\ref{eqn Kc}) it only
matters the number of neighbors on each site and, in some sense, the common
neighbors on the triangular lattice is counted twice.
The price paid, however, for such improvements in rather small systems 
taking the EFRG procedure clearly reflects 
the more elaborate calculations needed to obtaining the order parameters.
As a result, contrary to the MFRG, this drastically limits the size of
the employed clusters in such a way that the bigger system used
in literature is $N=4$.

Besides  pure and diluted Ising models 
\cite{li2,benayad2,fitti,ilkovic3,serkov,douglas3,douglasf} 
the EFRG has also been applied
in the study of the Ashkin-Teller model \cite{benayad2}, surface criticality
in semi-infinite Ising ferromagnets \cite{ilkovic4}, classical XY
and Heisenberg models \cite{ricar2b}, $O(n)$ vector ferromagnetic and
antiferromagnetic models \cite{ricar5}, the transverse Ising
model \cite{jiang3,jiang4}, and quantum spin-1/2 anisotropic Heisenberg
models \cite{ricar6,ricar7}. The same qualitative results have been achieved
in all applications.

\subsection{Finite Size Scaling Renormalization Group (FSSRG)}

The finite size scaling renormalization group (FSSRG) has been originally 
proposed by de Oliveira \cite{oliveira,oliveiraa} to treat Hamiltonian systems
having two energy terms. The main idea of the method is to consider
quantities having zero anomalous dimension, i.e., $\phi =0$ in equation
(\ref{eqn 3}). For the simple Ising model in an external magnetic field defined
in Eq. (\ref{eqn 0}) such quantities can be given by
\begin{equation} 
Q(K,H)=\left<sign\left({{1\over N}{\sum_i^NS_i}}\right)\right>~,
\label{eqn 58} 
\end{equation}
and
\begin{equation} 
R(K,H)=\left<sign\left({{1\over N_{s_1}}{\sum_i^{N_{s_1}}S_i}}\right)
sign\left({{1\over N_{s_2}}{\sum_i^{N_{s_2}}S_i}}\right)\right>~,
\label{eqn 59} 
\end{equation}
where $N=L^d$ is the number of spins on a d-dimensional lattice and 
$sign(x)= -1, 0, 1$, whether $x<0, x=0, x>0$, respectively. $s_1$ and
$s_2$ are two parallel surfaces with $N_{s_1}=N_{s_2}$ and separated 
by a distance $L/2$ apart when taking periodic boundary conditions. 
For finite lattices with open boundary conditions the
surfaces $s_1$ and $s_2$ can be taken as being the top and the bottom
hypersurfaces of the corresponding  hypercube, respectively. 

Let us consider first the quantity $Q$ and see that it has zero
anomalous dimension for an infinite system. While
the magnetization $m=\left<\left({{1\over N}{\sum_i^NS_i}}\right)\right>$
is zero for $T>T_c$ and behaves as
\begin{equation}
m\approx \epsilon ^\beta~,
\label{eqn 60}
\end{equation}
when $H\rightarrow 0^+$ and $T\rightarrow T_c^-$, the quantity $Q=+1$ for 
$T<T_c$ and is zero 
above $T_c$ meaning that $Q= \epsilon ^0$ and, from Eq. (\ref{eqn 0c}), 
$\phi =0$. Regarding now the second quantity $R$ we have that the probability
of finding ${{1\over N_{s_1}}{\sum_i^{N_{s_1}}S_i}}$ and
${{1\over N_{s_2}}{\sum_i^{N_{s_2}}S_i}}$ with the same sign is 
greater than finding them with opposite signs since, according to the weak
version of the Griffiths inequality \cite{ruelle}, one has 
$\left< S_iS_j \right> \ge 0$  for any
pair of spins on the lattice. Thus, for the same reasoning, we also have
a zero anomalous exponent for $R$.

By taking two finite systems of sizes $N,N^\prime$ and computing the
above quantities for both clusters one then gets the renormalization recursion 
relations
\begin{equation}
Q_{N^\prime}(K^\prime,H^\prime)=Q_N(K,H)~, 
\label{eqn 61} 
\end{equation}
and
\begin{equation}
R_{N^\prime}(K^\prime,H^\prime)=R_N(K,H)~. 
\label{eqn 62} 
\end{equation}

From equations above the complete renormalization flow in the $K-H$ plane can be
exploited. It turns out that for the Ising model the FSSRG gives the
quantitative correct behavior as $N>N^\prime \rightarrow \infty$
for all the fixed points 
$(K\rightarrow \infty ,H=0)$,  $(K=0,H=0)$  and  the Ising critical 
point $(K_c,H=0)$, while expected qualitative behavior is achieved
for finite systems. For instance, 
taking $N=4$ and $N^\prime =2$, which are the smallest possible systems
allowed by this approach, one readily gets
\begin{equation}
{\sinh (2H^\prime ) \over {Z_2}}={e^{4K}\sinh (4H)+4\sinh (2H) \over {Z_4}}~,
\label{eqn 63}
\end{equation}
\begin{equation}
{\cosh (2H^\prime)-e^{-2K^\prime} \over {Z_2}}=
{e^{4K}\cosh (4K) -1 \over {Z_4}}~,
\label{eqn 64}
\end{equation} 
where
\begin{equation}
Z_2=\cosh 2H^\prime + e^{-2K^\prime }~,
\label{eqn 65}
\end{equation}
\begin{equation}
Z_4=2+e^{-4K}+4\cosh (2H) +e^{4K}\cosh (4H)~,
\label{eqn 66}
\end{equation}
from which one obtains the values quoted in Table \ref{tab 7} for the Ising
critical point. Results from analytical calculations on bigger lattices 
are also listed in that table. 

However, contrary to the previous approaches,
the FSSRG method has the great advantage of allowing Monte
Carlo simulations for bigger finite systems in order to obtain
the desired quantities $Q$ and $R$ . Such Monte Carlo simulations are not so
easily implemented and has not yet been done in the mean field
like renormalization group treatments discussed herein (apart from a 
theory of mean field Monte Carlo simulation proposed by Netz and  Berker  in a
different context \cite{netz1,netz2,henriques}). 
These results are also shown in Table
\ref{tab 7} and, even taking into 
account the rather small computational effort, they are quite close to the 
exact ones. The FSSRG can also be used to treat other systems
with different kinds of energy and coupling constants
by conveniently choosing the quantities $Q$ and $R$ and even, 
in principle, be extended to study more than two parameter Hamiltonians,
such as flux diagrams in three dimensions.

Apart from the pure Ising model \cite{oliveira,oliveiraa,oliveirab}, the FSSRG
has also been applied to the diluted Ising model \cite{neto},
the spin-1 and spin-3/2  Blume-Capel model \cite{suzana}, and
the q-state Potts model \cite{stauffer}. An extension to the
dynamic critical behavior (for short and long times)
of the Ising \cite{mozart1} and Potts model \cite{mozart2}
has also been done through Monte Carlo simulations on finite systems. 
In this case, quite good accurate results for the dynamic critical exponents
$z$ have been achieved.

It should also be mentioned that the FSSRG approach is identical to the
phenomenological renormalization group  proposed by 
Nightingale \cite{nightingale} when
in the former one considers clusters consisting of infinite strips
with finite width \cite{pla10}. In order to see this equivalence let us
take a strip with finite width $L$ and length $n_L\rightarrow \infty$ at zero
external field (for simplicity). Correspondingly, the transfer matrix will 
be of order
$2^L\times 2^L$. In this case the quantity $Q_L$, computed
from the transfer matrix, is zero and one gets
$Q_{L^\prime}(K^\prime,H^\prime=0)=Q_L(K,H=0)=0$ for any values of
$L$ and $L^\prime <L$. On the other hand, the quantity $R_L$
is just a correlation function between the states of two columns separated by a
distance $n_L/2$ apart, which can be written as \cite{dombcl}
\begin{equation}
R_L(K)=A_L\left({{\lambda _L^< \over {\lambda _L^>}}}\right)^{n_L/2}~,
\label{eqn 67}
\end{equation}
where $A_L$ is a function of $K$ but independent of $n_L$ and $\lambda _L^>$ and
$\lambda _L^<$ are the greatest and second greatest eigenvalues of the
corresponding transfer matrix, respectively. For two strips with 
$L^\prime <L$ one  then has
\begin{equation}
A_{L^\prime}\left({{\lambda _{L^\prime}^< \over {\lambda _{L^\prime}^>}}}\right)
^{n_{L^\prime}/2}=
A_L\left({{\lambda _L^< \over {\lambda _L^>}}}\right)^{n_L/2}~,
\label{eqn 68}
\end{equation}
where $n_L,n_{L^\prime}\rightarrow \infty$. Taking now the logarithm of both
sides of the above equation, dividing by $n_{L^\prime}/2$ and recalling
that the scaling factor $\ell =L/L^\prime$ can also be obtained from
\begin{equation}
\ell = \left({N \over {N^\prime}}\right)^{1/d} =
\left({n_LL^{d-1} \over {n_{L^\prime}{L^\prime}^{d-1}}}\right)^{1/d}=
\left({L \over {L^\prime}}\right)~,
\label{eqn 69}
\end{equation}
(from which we have ${n_L \over {n_{L^\prime}}}={L \over {L^\prime}}$) 
we finally get
\begin{equation}
L^\prime \ln \left({{\lambda _{L^\prime}^< \over {\lambda _{L^\prime}^>}}}\right)=
L \ln \left({{\lambda _L^< \over {\lambda _L^>}}}\right)~,
\label{eqn 70}
\end{equation}
which is the same as Eq. (\ref{eqn 5}) since 
$\xi _L^{-1} = \ln \left({{\lambda _L^< \over {\lambda _L^>}}}\right)$
(the same for $L^\prime$).

It is interesting that two completely different phenomenological 
renormalization group procedures have the same (quite good)
results for the properties of one parameter critical models. 
One can think of the FSSRG as a generalization of the original 
Nightingale renormalization group for completely finite
systems (a fact that, in some way, explains the accuracy of the FSSRG 
critical values, specially in the Ising case). On the other hand, flux diagrams
for two-dimensional parameter Hamiltonians
(for instance, the Blume-Capel model \cite{luizdarcy}) taking infinite strips 
within FSSRG can be readily obtained (which is not possible from the
Nightingale procedure).

Another real space renormalization
group approach preserving the two spin correlation function in finite clusters
proposed by Tsallis et al \cite{tsallis0,tsallis} has also been shown to be
connected to the Nightingale procedure. However, it seems rather difficult to
implement the calculations by taking infinite self-dual clusters in the former
method.

\section{ Final Remarks}

In the above discussion we have been mainly concerned with
renormalization group procedures based on finite size scaling
hypothesis having a relatively wide application in rather
different statistical mechanics problems.
There are, however, some other renormalization group approaches based on 
finite lattice calculations. One of them, still close in spirit to
the ones we have been described so far, is the new mean field 
renormalization group \cite{jaff} transformation (NMFRG).  
To briefly illustrate this
method let us recall Eq. (\ref{eqn 5a}) in the mean field approximation
$m_N=b$. Close to the mean field transition $T_c^N$ the order parameter can
be expanded as 
\begin{equation}
m_N=F_N(K)\epsilon ^{\beta ^*}_N + G_N(K)H\epsilon ^{-\gamma ^*}_N ~,
\label{eqn 71}
\end{equation}
where $\epsilon = (K-K_c^N)/K$, $K_c^N$ is the mean field critical temperature and
$\beta ^*$ and $\gamma ^*$ are the usual mean field exponents
($\beta ^*=1/2$ and $\gamma ^*=1$). At $K=K_c^N$ the
functions $F_N(K)$ and $G_N(K)$ are called anomaly coefficients, 
introduced by Suzuki \cite{suzuki2,suzuki3} and from which non-classical 
exponents are also achieved (the so-called coherent anomaly method - CAM). 
In the present renormalization group context one expands the order parameter
for two different clusters $N^\prime < N$ and assumes a scaling relation
of the form given in Eq. (\ref{eqn 6}) for the approximate order parameters 
and the same relation for the quantities $\epsilon ^{\beta ^*}_N$ and
$\epsilon ^{\beta ^*}_{N^\prime}$, i.e., $\epsilon ^{\beta ^*}_{N^\prime}=
\ell ^{d-y_H}\epsilon ^{\beta ^*}_N$. So, in the same way as previously done for the
MFRG one then gets
\begin{equation}
F_{N^\prime}(K^\prime)=F_N(K)~,
\label{eqn 72}
\end{equation}
from which the critical fixed point $K_c=K^\prime=K$ is obtained, as well as
the thermal exponent $\nu$, whereas the corresponding magnetic exponent is given by
\begin{equation}
\ell ^{y_H}=\left({\epsilon_{N^\prime} \over {\epsilon_N}}\right)^{\beta^*+\gamma^*}
{G_N(K_c) \over {G_{N^\prime}(K_c)}}~.
\label{eqn 73}
\end{equation}
It is apparent that although the anomaly  coefficients are based on the
classical values $\beta ^*$ and $\gamma ^*$, non-classical exponents
are extracted from the linearization of the recursion relation 
(\ref{eqn 72}) around the
fixed point. This approach gives quite good results
for the Ising model as well as for geometrical phase transitions \cite{jaff},
and can also be extended by using the effective field theory through
the use of Callen-Suzuki identity \cite{coelho}. It is also verified that
the larger the value of $N$ and $N^\prime$, the better the approximation since
more fluctuations are included in the mean field calculations within each
cluster.

Just to mention, another approach, used in a different context, is the 
density matrix renormalization group
\cite{white1,white2} which gives accurate numerical results in
studying one-dimensional quantum lattice models \cite{noack},
the two-dimensional Ising model \cite{nishino} and strongly correlated
electron systems \cite{white3}.
Nevertheless, it is clear that other phenomenological renormalization 
group proposals can still be done by exploiting relation (\ref{eqn 3}) 
for quantities where $\phi$ is known in the infinite system. 

Concerning the present schemes there are still some open problems 
whose study would be natural extensions
of the methods discussed herein. Among them, for instance, we have:

{(i)} there is no unified approach (as in the MFRG case) to the SBMFRG (and its
corner field partner) in order to get complete flux diagrams for models
with more than one Hamiltonian parameter space. In fact, just for the Ashkin-Teller 
model a flow diagram in the Hamiltonian parameter space has been obtained through
the SBMFRG \cite{oliveira2};

{(ii)} there is also a lack of a SBMFRG scheme to treat  critical
dynamics by considering three clusters. Only the two cluster approach has
been implemented in the MFRG to dynamics;

{(iii)} it would be interesting to extend the EFRG procedure to account
for obtaining flow diagrams;

{(iv)} flow diagrams in more than two Hamiltonian parameter space has not yet
been considered by using neither the MFRG nor the FSSRG approach;

{(v)} finally, Table \ref{tab 8} shows roughly a picture of the models
treated by the MFRG approach presented in sections III-V. It is also
presented the applications of the other procedures discussed herein.
So, besides new systems to be treated and studied from the present
approaches, all the lacunae in Table \ref{tab 8} are straight 
extensions to be done on this subject. Perhaps, some readers will find
it easy to readily fill in some gaps.

\acknowledgements  

{The authors would like to thank  J. Ricardo
de Souza,  P. M. C. de Oliveira, J. Kamphorst Leal da Silva and
J. G. Moreira for valuable discussions.
Financial support from CNPq, FINEP, CAPES and FAPEMIG (Brazilian Agencies)
is also gratefully acknowledged.}
 
\begin{table}\caption  ~Critical values for the $d=2$ Ising model 
                       on a square lattice according
                       to the ordinary MFRG (two clusters) and SBMFRG
                       (three clusters). The two entries for the critical
                       indices in the SBMFRG correspond to the values given 
                       from equations (\ref{eqn 16a}) and (\ref{eqn 16b}),
                       respectively \tablenotemark[1].
\begin{tabular}{cccccc}
$  $  $ {\bf MFRG} $  $  $  \\
\tableline
$N,N^\prime$ &$ K_c$ &$ \nu $ &$ y_H$& $y_{HS}$&$\nu y_H$\\
\tableline
~~4,~1  & 0.361   & 1.45  & 1.50  &  &  2.17 \\
~~9,~4  & 0.381   & 1.28  & 1.57  &  &  2.01 \\
~16,~9  & 0.393   & 1.22  & 1.60  &  &  1.95 \\
~25,16  & 0.401   & 1.19  & 1.62  &  &  1.93 \\
\tableline
$  $  $ {\bf SBMFRG} $   $  $  \\
\tableline
$N,N^\prime,N^{\prime\prime}$ &$ K_c$ &$ \nu $ &$ y_H$& $y_{HS}$&$\nu y_H$\\
\tableline
~9,~4,~1  & 0.413   & 1.23-1.37  & 1.63-1.56 & 0.50-0.56 & 2.00\\
16,~9,~4  & 0.425   & 1.16-1.22  & 1.68-1.65 & 0.49-0.52 & 1.95\\
25,16,~9  & 0.430   & 1.13-1.16  & 1.71-1.69 & 0.49-0.51 & 1.92\\
36,25,16  & 0.433   & 1.10-1.12  & 1.73-1.72 & 0.49-0.50 & 1.91\\
\tableline
exact \tablenotemark[2] & 0.441  &  1 & 1.875 & 0.5 \tablenotemark[3] & 1.875 \\
\end{tabular}

\tablenotemark[1]{Data from Refs. \cite{indekeu}
                       and \cite{croes} for hypercubic finite clusters.}
                       
\tablenotemark[2]{Refs. \cite{onsager,baxter}.}

\tablenotemark[3]{Ref. \cite{binder}.}                     
\label{tab 1}\end{table}

\begin{table}\caption ~The same as Table I for the $d=3$ simple cubic Ising model.
                      For hypercubic clusters one 
                      has $N=L^3$, and for  rectangular prisms  
                      $N=L(L-1)^2$ or $N=L^2(L-1)$ \tablenotemark[1].
\begin{tabular}{cccccc}
$  $  $ {\bf MFRG} $  $  $  \\
\tableline
$N,N^\prime$ &$ K_c$ &$ \nu $ &$ y_H$& $y_{HS}$&$\nu y_H$\\
\tableline
~~2,~1  & 0.203   & 1.50  & 1.89  &  &  2.84 \\
~~4,~1  & 0.207   & 1.22  & 2.00  &  &  2.44\\
~~9,~4  & 0.212   & 1.05  & 2.08  &  &  2.18\\
\tableline
$  $  $ {\bf SBMFRG} $   $  $  \\
\tableline
$N,N^\prime,N^{\prime\prime}$ &$ K_c$ &$ \nu $ &$ y_H$& $y_{HS}$&$\nu y_H$\\
\tableline
~4,~2,~1  & 0.212   & 1.23-1.49  & 2.00-1.91 & 1.03-1.12 & 2.85\\
~8,~4,~2  & 0.215   & 0.96-1.22  & 2.15-2.00 & 0.90-1.05 & 2.44\\
12,~8,~4  & 0.201   & 1.32-1.03  & 1.95-2.09 & 0.99-0.85 & 2.15\\
18,12,~8  & 0.216   & 1.04-1.19  & 2.09-2.01 & 0.95-1.03 & 2.39\\
27,18,12  & 0.217   & 0.88-0.99  & 2.20-2.09 & 0.86-0.97 & 2.07\\
\tableline
other methods& 0.222 \tablenotemark[2] &  0.63 \tablenotemark[3]     & 2.48 
\tablenotemark[3] & 0.8 \tablenotemark[4] & 1.56 \\
\end{tabular}

\tablenotemark[1]{Data from Refs. \cite{indekeu}
                       and \cite{croes}.}
                       
\tablenotemark[2]{Ref. \cite{domb1,landau3d,bloete1,bloete2}.}

\tablenotemark[3]{Ref. \cite{landau3d,bloete1,bloete2,guillou}.}

\tablenotemark[4]{Ref. \cite{binder}.}
                       
\label{tab 2}\end{table}

\begin{table}\caption ~The same as Table I for the critical values of the $d=2$ 
                       site percolation problem.
\begin{tabular}{cccc}
$  $  $ {\bf MFRG} $  $  $  \\
\tableline
$N,N^\prime$ &$ p_c$ &$ \nu_p $ & $y_{HS}^p$ \\
\tableline
~~2,~1  & 1/3     & 1.56  &   \\
~~4,~2  & 0.427   & 1.52  &   \\
~~9,~4  & 0.443   & 1.47  &   \\
~16,~9  & 0.473   & 1.43  &   \\
\tableline
$  $  $ {\bf SBMFRG} $   $  $  \\
\tableline
$N,N^\prime,N^{\prime\prime}$ &$ p_c$ &$ \nu_p $ & $y_{HS}^p$\\
\tableline
~9,~4,~1  & 0.602   &  1.25-1.35 & 0.72-0.69 \\
16,~9,~4  & 0.585   &  1.35-1.35 & 0.66-0.65 \\
\tableline
other methods& 0.593 \tablenotemark[1] &  1.33 \tablenotemark[2]     &  --  \\
\end{tabular}

\tablenotemark[1]{Ref. \cite{stauffer0}.}
                       
\tablenotemark[2]{Ref. \cite{essam}.}
                       
\label{tab 2a}\end{table}

\begin{table}\caption ~Critical values for the $d=2$ Ising model 
                       on a square lattice according
                       to the SBMFRG including the corner magnetic exponent
                       $y_{HC}$ \tablenotemark[1]. 
\begin{tabular}{cccccc}
$N,N^\prime,N^{\prime\prime}$ &$ K_c$ &$ \nu $ &$ y_H$& $-y_{HC}$\\
\tableline
16,~9,~4  & 0.469   & 1.08-1.13 & 1.78-1.72 & 0.57-0.51 \\
25,16,~9  & 0.460   & 1.07-1.10 & 1.80-1.76 & 0.63-0.59 \\
36,25,16  & 0.454   & 1.06-1.07 & 1.81-1.78 & 0.68-0.65 \\
\tableline
 exact \tablenotemark[2]& 0.441  & 1   & 1.875  & 1 
 \tablenotemark[3]  \\
\end{tabular}

\tablenotemark[1]{Data from Ref. \cite{croes} applying clusters 
                       in which the corner exponent can be defined.}
                       
\tablenotemark[2]{Ref. \cite{onsager,baxter}.}

\tablenotemark[3]{Ref. \cite{cardy}.}                      

\label{tab 3}\end{table} 

\begin{table}\caption{ Critical temperature and dynamic critical exponent
                       for the Ising model according to the DMFRG.}                      
\begin{tabular}{||c|c|c||c|c|c||} \hline
\multicolumn{3}{||l|}{d=2} & 
\multicolumn{3}{l|}{d=3} \\  \hline
$N,N^\prime$  &  $K_{c}$  &  z    &  $N,N^\prime$  &  $K_{c}$  &
  z  \\  \cline
{1 - 6}
 2,1  &  0.347  & 1.17   &  2,1  &  0.203  &   0.97  \\
 4,1  &  0.361  & 1.39   &  4,2  &  0.207  &   1.15  \\
 4,2  &  0.370  & 1.60   &  8,2  &  0.207  &   1.32  \\
 9,4  &  0.381  & 2.13   &  8,4  &  0.209  &   1.49  \\  \cline
{1 - 6}
    &   $0.441 \tablenotemark[1]$  &  $2.2\pm 0.2 \tablenotemark[2]$   &    &  
$0.222 \tablenotemark[3]$ &  
$2.0 \tablenotemark[4]$  \\  \hline
\end{tabular}

\tablenotemark[1]{Refs. \cite{onsager,baxter}.}

\tablenotemark[2]{Ref. \cite{Jan}.}

\tablenotemark[3]{Ref. \cite{domb1,landau3d,bloete1,bloete2}.}

\tablenotemark[4]{Ref. \cite{Dominicis}.}

\label{tab 4}\end{table}

\begin{table}\caption{Surface dynamic critical exponent $z_S$ and critical
                      surface coupling $\rho _c$ for the three-dimensional
                      Ising model according to the DMFRG.}
\begin{tabular}{ccc}
$N,N^\prime$ & $z_S$ & $ \rho _c$ \hfill \\
\tableline
~~2,~1  & 1.35   & 1.35  \hfill \\
~~4,~2  & 1.88   & 1.48  \hfill \\
\tableline
 Monte Carlo  \tablenotemark[1]     & ---  &  1.52   \hfill \\
\end{tabular}

\tablenotemark[1]{Ref. \cite{Landau}.}

\label{tab 5}\end{table} 

\begin{table}\caption{ Critical temperature and thermal critical exponent
                       for the Ising model according to the EFRG. It is
                       quoted in parenthesis the values from MFRG for
                       comparison.}                      
\begin{tabular}{||c|c|c||c|c|c||} \hline
\multicolumn{3}{||l|}{d=2} & 
\multicolumn{3}{l|}{d=3} \\  \hline
$N,N^\prime$  &  $K_{c}$  & $\nu$    &  $N,N^\prime$  &  $K_{c}$  &
 $\nu$  \\  \cline
{1 - 6}
 2,1 &  0.358 (0.347)  & 1.39 (1.67)  &  2,1 & 0.206 (0.203) & 1.37 (1.50) \\
 4,1 &  0.379 (0.361)  & 1.17 (1.45)  &  4,1 & 0.207 (0.207) & 1.24 (1.22) \\
 4,2 &  0.371 (0.370)  & 1.01 (1.28)  &  4,2 & 0.208 (0.207) & 1.14 (1.27) \\
 \cline
{1 - 6}
    &   $0.441\tablenotemark[1]$  &  1 \tablenotemark[1]  &    &  
$0.222 \tablenotemark[2]$ &  0.63 \tablenotemark[3]  \\  \hline
\end{tabular}

\tablenotemark[1]{Refs. \cite{onsager,baxter}.}

\tablenotemark[2]{Refs. \cite{domb1,landau3d,bloete1,bloete2}.}

\tablenotemark[3]{Refs. \cite{landau3d,bloete1,bloete2,guillou}.}

\label{tab 6}\end{table}

\begin{table}\caption  ~Critical temperature and thermal critical exponent
                       for the Ising model according to the FSSRG. The figures
                       for $N=4096$ and $N^\prime =1024$ are from Monte
                       Carlo simulations with the parenthesis indicating
                       the first uncertain digit, according to statistical 
                       fluctuations \tablenotemark[1].                       
\begin{tabular}{||c|c|c||c|c|c||} \hline
\multicolumn{3}{||l|}{d=2} & 
\multicolumn{3}{l|}{d=3} \\  \hline
$N,N^\prime$  &  $K_{c}$  & $\nu$    &  $N,N^\prime$  &  $K_{c}$  &
 $\nu$  \\  \cline
{1 - 6}
 4,2       &  0.473     & 0.905    &  8,2 & 0.225 & 1.052 \\
 16,4      &  0.432     & 1.053    &      &       &       \\
 4096,1024 &  0.440(8)  & 1.00(0)  &      &       &       \\
 \cline
{1 - 6}
    &   $0.4407 \tablenotemark[2]$  &  1 \tablenotemark[2]  &    &  
$0.222 \tablenotemark[3]$ &  0.63 \tablenotemark[4]  \\  \hline
\end{tabular}

\tablenotemark[1]{Data from reference \cite{oliveira}.}

\tablenotemark[2]{Refs. \cite{onsager,baxter}.}

\tablenotemark[3]{Refs. \cite{domb1,landau3d,bloete1,bloete2}.}

\tablenotemark[4]{Refs. \cite{landau3d,bloete1,bloete2,guillou}.}

\label{tab 7}\end{table}

\begin{table}\caption  ~Some models studied by means of the methods
                        described in the text (indicated by ``x").
\begin{tabular}{ccccc}
 model & MFRG & EFRG & FSSRG & NMFRG \\
\tableline
 {\bf geometrical} & x &  & &   \\
\tableline
 {\bf Classical (static)} &  &  & &   \\
\tableline
pure Ising like     &  x  & x & x & x  \\
random  Ising like  &  x  & x & x & x  \\
surface critical    &  x  & x &   &   \\
spin-glass          &  x  &   &   &    \\
random-fields       &  x  &   &   &   \\
XY and Heisenberg   &  x  & x &   &   \\
Ashkin-Teller       &  x  & x &   &   \\
ANNNI               &  x  &   &   &   \\
mixed spin          &  x  &   &   &   \\
Z(q) symmetric spin &  x  &   &   &   \\
random D-vector     &  x  &   &   &   \\
$q$-state clock spin-glass&  x  &   &   &   \\
Potts               &  x  &   & x &   \\
Blume-Capel         &  x  &   & x &    \\
lattice gas         &  x  &   &   &    \\
liquid crystal      &  x  &   &   &    \\
Tsallis Statistics  &  x  &   &   &   \\
\tableline
 {\bf Quantum (static)} &  &  & &   \\
\tableline
pure and random transverse Ising        & x  & x &  & \\
anisotropic Heisenberg                  & x  & x &  & \\
random-mixture                          & x  &   &  & \\
Dzyaloshinskii-Moryia                   & x  &   &  & \\
granular superconductors                &  x  &   &  &    \\
\tableline
 {\bf Dynamic} &  &  & &   \\
\tableline
Ising like         &  x  &   & x & \\
Potts              &     &   & x & \\
contact-process    &  x  &   &   & \\
surface critical   &  x  &   &   & \\
quantum spin       &  x  &   &   & \\
cellular automata  &  x  &   &   &    \\   
\end{tabular}                    
\label{tab 8}\end{table}

\begin{figure}
\caption{ (a) and (b) show schematically the smallest clusters for
 hypercubic lattices and (c) and (d) the same for the directed self
 avoiding random walk (DSARW) in two dimensions. Full circles represent sites
 belonging to the cluster itself and open circles the corresponding
 surrounding sites (mean field sites).}
\label{Fig1}
\end{figure}

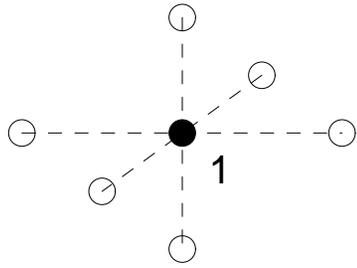
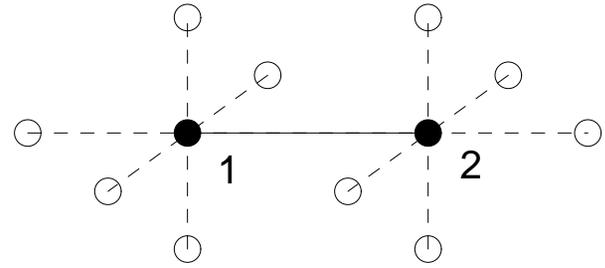
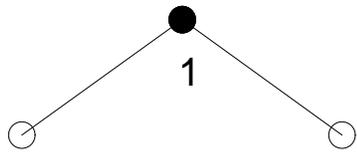
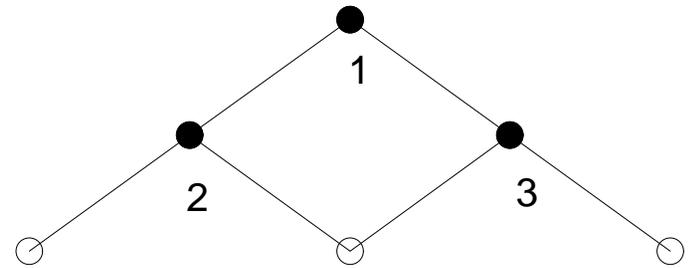
\end{document}